\documentclass[10pt,journal,compsoc]{IEEEtran}
\pdfoutput=1

\ifCLASSOPTIONcompsoc
  \usepackage[nocompress]{cite}
\else
  \usepackage{cite}
\fi
\ifCLASSINFOpdf
\else
\fi
\usepackage{booktabs} % For formal tables
\usepackage{mathrsfs}
\usepackage{graphicx}
\usepackage{subfigure}
\usepackage[justification=centering]{caption}
\usepackage{url}
\usepackage{enumitem}
\usepackage{epsfig,endnotes}
\usepackage{hyperref}
\usepackage{float}
\usepackage{graphicx}
\usepackage{lscape}
\usepackage{algorithm}
\usepackage[noend]{algpseudocode}
\usepackage{commath}
\usepackage{color}
\usepackage{amsfonts}
\usepackage{amsmath,amssymb}
\usepackage{gensymb}

\newcommand{\name}{ReCoil}

\newcommand{\tabincell}[2]{\begin{tabular}{@{}#1@{}}#2\end{tabular}}

\begin{document}
\title{You foot the bill! Attacking NFC with passive relays}

\author{Yuyi~Sun,~\IEEEmembership{Student Member,~IEEE,}
Swarun~Kumar,~\IEEEmembership{Member,~IEEE,}
Shibo~He,~\IEEEmembership{Senior Member,~IEEE,}
Jiming~Chen,~\IEEEmembership{Fellow,~IEEE,}
and~Zhiguo~Shi,~\IEEEmembership{Senior Member,~IEEE}
        %        <-this % stops a space
\IEEEcompsocitemizethanks{\IEEEcompsocthanksitem Y. Sun, S. He, J. Chen and Z. Shi are with State Key Laboratory of Industrial Control Technology, Zhejiang University, Hangzhou, 310027, China, E-mail: yuyisun@zju.edu.cn, s18he@iipc.zju.edu.cn, cjm@zju.edu.cn, shizg@zju.edu.cn.\protect\\

\IEEEcompsocthanksitem S. Kumar is with the Department of Electrical and Computer Engineering, Carnegie Mellon University, Pittsburgh, PA, 15213, USA, E-mail: swarun@cmu.edu.}% <-this % stops an unwanted space
}
\IEEEtitleabstractindextext{%
\begin{abstract}
Imagine when you line up in a store, the person in front of you can make you pay her bill by using a passive wearable device that forces a scan of your credit card without your awareness. An important assumption of today's Near-field Communication (NFC) enabled cards is the limited communication range between the commercial reader and the NFC cards -- a distance below 5~cm. Previous approaches to attacking this assumption effectively use mobile phones and active relays to enlarge the communication range, in order to attack the NFC cards. However, these approaches require a power supply at the adversary side, and can be easily localized when mobile phones or active relays transmit  NFC signals.

We propose \name, a system that uses wearable passive relays to attack NFC cards by expanding the communication range to 49.6 centimeters, a ten-fold improvement over its intended commercial distance. \name\ is a magnetically coupled resonant wireless power transfer system, which optimizes the energy transfer by searching the optimal geometry parameters. Specifically, we first narrow down the feasible area reasonably and design the \name-Ant Colony Algorithm such that the relays absorb the maximum energy from the reader. In order to reroute the signal to pass over the surface of human body, we then design a half waist band by carefully analyzing the impact of the distance and orientation between two coils on the mutual inductance. Then, three more coils are added to the system to keep enlarging the communication range. Finally, extensive experiment results validate our analysis, showing that our passive relays composed of common copper wires and tunable capacitors expand the range of NFC attacks to 49.6 centimeters.
\end{abstract}

\begin{IEEEkeywords}
Near-field communication, passive relay, magnetically coupled resonant wireless power transfer, attack.
\end{IEEEkeywords}}

\maketitle

\IEEEdisplaynontitleabstractindextext

% \fi
\IEEEpeerreviewmaketitle

\IEEEraisesectionheading{\section{Introduction}\label{Introduction}}

%The \textit{proceedings} are the records of a conference.\footnote{This
%  is a footnote}  ACM seeks
%to give these conference by-products a uniform, high-quality
%appearance.  To do this, ACM has some rigid requirements for the
%format of the proceedings documents: there is a specified format
%(balanced double columns), a specified set of fonts (Arial or
%Helvetica and Times Roman) in certain specified sizes, a specified
%live area, centered on the page, specified size of margins, specified
%column width and gutter size.

\IEEEPARstart{I}{magine} you are waiting in line at your favorite store, but unbeknownst to you, you just paid for the person in front of you. \textit{And}, the person in front of you achieved this without any active electronics, making her attack virtually untraceable and (technically) FCC compliant~\cite{nelson2017red}! Recent years have witnessed a proliferation of Near-field Communication (NFC) for secure and efficient communication in the Internet of Things. NFC works at the frequency 13.56~MHz defined by the standard ISO/IEC 14443 \cite{ISOIEC14443-42018} and allows inexpensive battery-free tags to communicate wirelessly with powered readers over extremely short range (5 centimeters at most). In principle, NFC limits the maximum communication distance between the tag and reader to extremely close distances -- an attractive security guarantee~\cite{rahman2017classification, dang2017large}. Indeed, NFC has seen wide adoption for secure battery-free systems such as payment cards and identity (ID) cards.

\begin{figure}%[htbp]
\begin{center}
\includegraphics[width=0.4\textwidth]{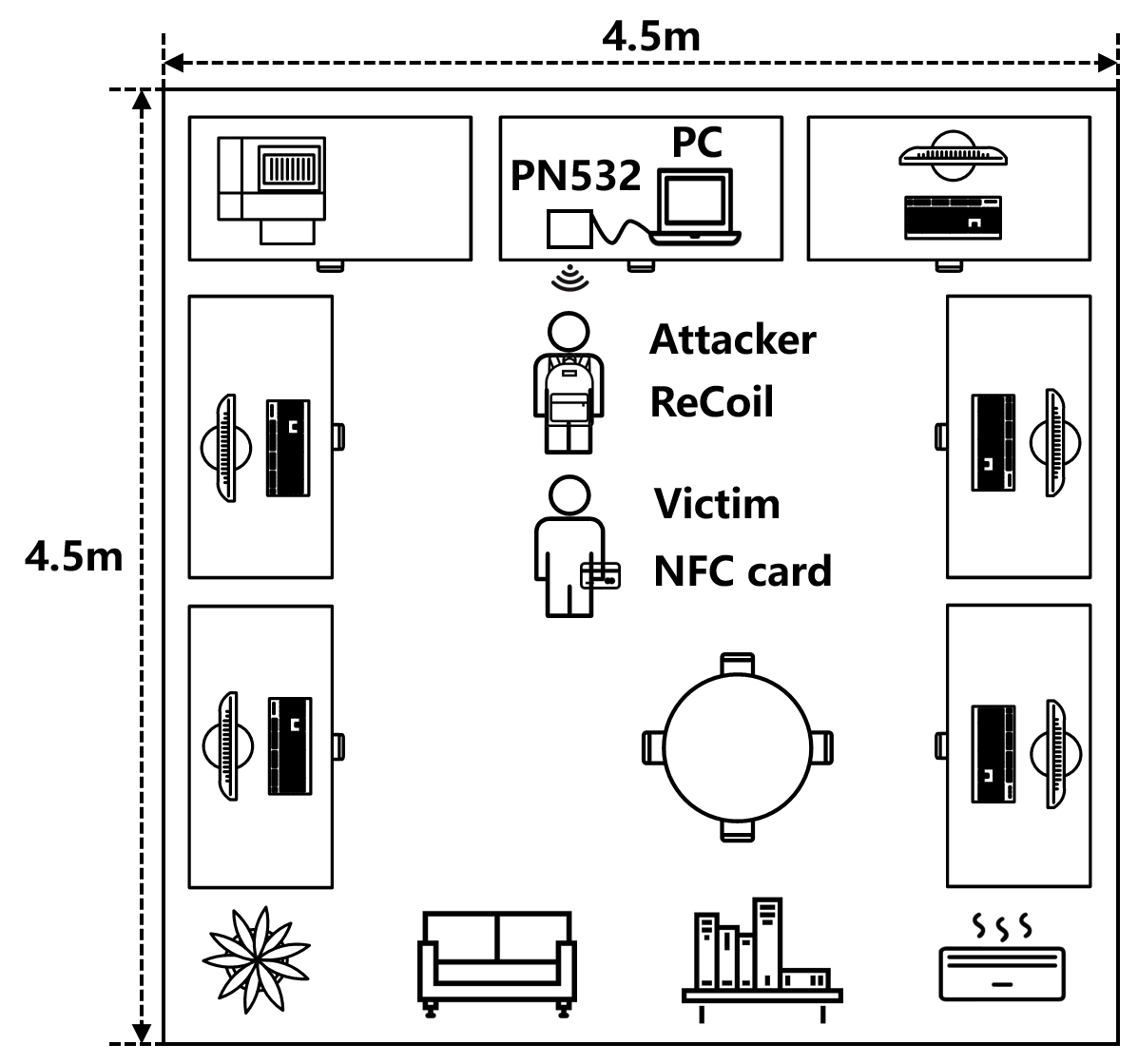}
\caption{The \name\ scenario.}
\label{fig_implement}
\end{center}
\end{figure}

%There has been much work done on NFC hacking based on smartphones \cite{bermejo2017steal, hur2017survey, jumic2017analysis}. However, using smartphones is more likely to be detected, which would always be more energy-consuming. Similarly, there has been some work done about using active relays to hack NFC cards \cite{wang2012implementation, forrester2018assessing}. In this paper, we build XXX, a system that can hack NFC access card using passive multiple-input-multiple-output (MIMO) coil relays. Specifically, we target NFC access cards, the basic and simple type without largely encrypted accounts or complicated raw data in the protocols, which can be hacked fast and frequently. The passive MIMO coil relays are made from copper wires and capacitors, which can be found easily. Then, we line up many relays in parallel as a waist band, using the mutual inductance to transmit magnetic power. This enables the magnetic field to form a closed circle, transmitting the signal of the reader more far away. Finally, with the help of X people wearing this kind of band, we can hack the access cards within one-meter distance and grasp the ID information.

However, recent studies have revealed several weaknesses of the NFC framework \cite{bermejo2017steal, hur2017survey, jumic2017analysis, francis2010practical, markantonakis2012practical}. Malicious software on cellphones can exploit bugs in the Android NFC Stack or browser to attack NFC Services. For instance, infected phones close to contactless cards can surreptitiously read their information. More recent novel work has demonstrated using active relays to attack NFC cards \cite{wang2012implementation, korak2014power, kfir2005picking}. A smartphone or a circuit board with Android SDK API can serve as a man-in-the-middle, that relay signals from the NFC tag to the reader, violating the proximity assumption NFC communications.
% However, active relays are relatively easy to be localized because of their signal transmissions~\cite{ma2013location} and the conspicuous latency in the exchange of packets between the card and reader~\cite{gurulian2017effectiveness}.

%Imagine a scenario in which you are standing in front of the counter at one of Starbucks' stores, after you order a latte, the people who stand just behind you or in the line just beside you would pay for your order without their knowledge. With our system, this will be possible.

% In this paper, we build \name\ -- to our knowledge, the first completely passive (battery-free) man-in-the-middle attack that extends the range of NFC. We allow NFC cards to be red at over ten times their normal range using a sequence of passive relays that are thin, light-weight and readily shielded by clothing.
% %In our scenario, we target NFC cards, the basic and simple type without largely encrypted accounts or complicated raw data in the protocols, which can be hacked fast and frequently.
% These relays effectively steer near-field signal beams between the transmitter-receiver pair across further distances by up to ten times their normal range. A detailed experimental evaluation on commercial NFC readers and cards shows that we can relay query signals and responses between them over distances of 49.6 centimeters between them, against the advertised range limit of 5 centimeters.

In this paper, we build \name\ -- that, to our knowledge, provides the first completely passive attack on NFC applications. Passive relays are adopted due to the following reasons: 1) our device is totally composed of copper wires and capacitors, making the system a low-cost and low-consumption one. 2) We design a wearable device which can be totally shielded by clothing or equipment that the attacker carries. Without power supply and active NFC reader, it is more possible for the attacker to attack the victim stealthily and secretly. 3) Since we do not revise the commercial protocols of NFC, our device is hard to detect. We allow NFC cards to be read at obover ten times their normal communication range using an array of passive relays that are thin, light-weight and small (readily shielded by clothing). These passive relays equipped with variable capacitors can steer near-field signal to provide maximized power to the NFC cards over further distances.
%In our scenario, we target NFC cards, the basic and simple type without largely encrypted accounts or complicated raw data in the protocols, which can be hacked fast and frequently.
A detailed experimental evaluation on commercial NFC readers and cards shows that we can relay query signals and responses between them over distances of 49.6 centimeters, under the environment shown as Fig.~\ref{fig_implement}, against the advertised range limit of 5 centimeters.

%However, achieving the goals above has two challenges: increasing the distance between the commercial reader and passive tag with passive coil relays, and eliminating the interference from human body. Firstly, without additional power supply, our system enlarges the communication range only by regulating the coil antenna to obtain better performance of the passive relays. This would be further complicated due to the distance and orientation among these coil relays because the inductance can be easily affected by the separation between two coils, making the resonance frequency change and the MIMO performance decrease. The other challenge is human body interference when putting the waist band made from passive coil relays on our bodies. The metallic materials on people's clothes, like the zippers or some metallic decorations, can effortlessly interfere the magnetic field, also changing the resonance frequency and decreasing the signal transmitted.

% Our core contribution is exploring the design space of the passive relays between the NFC card and reader. Our relay design is simple -- an array of coils composed of common copper wires and capacitors shielded by clothing. These coils change how the signals between the NFC reader and card couple, effectively directing energy over much further distances than originally designed. Our key objective is to therefore explore the design of this coil to maximize energy-transfer between the card and reader.

Our core contribution is providing a near-field passive solution to maximize the output power to the NFC cards. Since the NFC tag is a purely batteryless device,  \name\ fully relies on the energy delivered wirelessly by a nearby NFC reader. Without additional power supply, it enlarges the communication range only by judiciously regulating the geometry parameters of the coils to obtain an optimized performance of the passive relays. This is complicated due to the fact that the distance and orientation among these coil relays can easily affect the mutual inductance between any two adjacent coils, resulting in a changed resonance frequency and downgraded performance. Therefore, it is important to  optimize the design and geometry of the coils to maximize the energy transfer between the passive relays. \name\ achieves this by carefully studying the inductance of the copper coils, through extensive modulations, simulations and experimentation. We first study varying shapes of the coils and show that square coils have better performance than other antennas. We then formulate an optimization problem to maximize the output power and compute the optimal parameters of the passive coils. Since non-linear solvers  usually converge to sub-optimal local maxima, we design the \name-Ant Colony Algorithm (RACA) to randomly initiate the beginning parameters and find the optimal geometry parameters which maximize the output power.

% By repeating this progress, our algorithm finally converges to the optimal parameters.

% However, in the context of near-field communication, the passive relays couple with each other, which means that changing the phase of the signal relayed from one passive relay can also change the phase of the signals relayed by others. This means that that a modest change in phase at any one coil can radically change the effectiveness of beamforming across all coils. To make matters worse, the precise correlation in phase between coils is not deterministic and can change with the relative spacing, orientation and geometry of the passive relays.

% To resolve this dilemma, \name\ develops a gradient-based search algorithm to find the optimal sets of beamforming weights that can beam maximized power at furthest distance. At a high level, \name\ first considers the mutual influence of adjacent passive relays for different beamforming weights applied across passive relays. It then models energy distribution in the space for each set of beamforming weights. We uses a gradient-based approach to favor the beamforming weights deliver high energy at further distance. By repeating this progress, our algorithm finally converge at the optimal set of beamforming weights that can provide sufficient energy to the NFC card at furthest distance.

%Based on the online model eDesignSuite \cite{edesignsuite2018}, we circularly calculate the inductance of the rectangular coils, and use vector network analyzer (VNA) or LCR meter to verify this value and revise the shape.
A second challenge that \name\ addresses is to reroute the signal to pass over the surface of the human body because the signal is too weak to go through the human body directly. We build a Magnetically Coupled Resonant -  Wireless Power Transfer (MCR-WPT) system to mitigate this challenge.  Our design is a wearable waistband that surrounds the body and therefore mitigates its attenuation of the NFC signal when attacking the targetted NFC card. We add three more coils which are hidden in a backpack to keep the signal transmission. We then describe how the location of the coils can be optimized to maximize coupling between the NFC transmitter-receiver pair over extended distances. We further adapt the resonant frequency of the coils to the frequency of NFC signal by tuning the capacitor attached to the coils. A key challenge in ensuring the optimized energy transfer is the impact of objects between the relay and transmitter-received pair -- the body and clothing of the attacker. When the attack device is attached on human body, the resonated frequency can be affected by complicated environment. Thus, we first perform experiments to shield coils from metallic materials on clothes by building a Ferrite sheet and model its impact. Further, our results in multipath-rich settings show that our system can still work at complicated environment in a robust manner.

% \textbf{Limitations:} Though \name\ coils are designed to be relatively inflexible, minute shape changes of the array can cause the resonant frequency to easily change, impacting the performance of the relays. In addition, our system can be affected by all-metal objects, like mobile phones or keys. Besides, the commercial readers or the passive tags should be opposite to one of the coils, ensuring as much magnetic flux as possible passes through the coil, and the tag receive the largest power. Sec. \ref{discussion} discusses these limitations.

We implement \name\ as a light-weight multi-coil array that can be firmly worn on human bodies. We use Adafruit PN532 which has micropower as the commercial reader, making \name\ a typical power-limited system. AWG 18 copper wires and variable capacitors are used to create the coils and apply intended phase shift. Based on the analysis and models, we use the High Frequency Structure Simulator (HFSS) to simulate and validate the performance of the passive coil relays by changing the corresponding orientation and separation. Based on these simulations and experimentation, we design the optimal dimension, distance and orientation of coils, achieving the optimal communication range. The results from our testbed show:

\begin{enumerate}
\item a distance of 49.6~cm in accessing other people's NFC cards, ten times that of the existing NFC systems.
\item robustness when attacking the NFC cards under multipath over a distance of 48~cm.
\item successful attack over multiple NFC cards.
\item a practical deployment compatible with commercial NFC-based smartphones.
\end{enumerate}

\textbf{Contributions:} We propose a novel system  attacks the proximity-assumption of NFC using passive multi-coil relays. We design the optimal dimension, separation and orientation of these coils. Our system achieves extending the range of NFC cards successfully over distances of to 49.6~cm with negligible increase in latency. %With the help of this array, the anti-collision system can hack XX cards simultaneously.

The rest of the paper is organized as follows. In Sec.~\ref{related}, we discuss related work on attacking NFC, passive relays design and how to detect potential attacks. NFC principles and the MCR-WPT model are introduced in Sec. \ref{background}. Attack model, key challenges and solutions are explained in Sec.~\ref{systemmodel}. Sec.~\ref{systemdesign} proposes our system design, mainly focusing on our algorithm and coil design. We implement \name\ in Sec.~\ref{implementation} and conduct extensive evaluations in Sec.~\ref{evaluation}. Then, we discuss the limitations of our system and future work in Sec.~\ref{discussion}. Sec. \ref{conclusion} concludes the paper.

\section{Related Work}\label{related}

\textbf{Active Attacks in NFC:} Though NFC has limited communication range to guarantee the security of payments, there has been much work on active NFC relay hacking \cite{lee2012nfc}. Much novel work increased the NFC range by the relay attack using smartphones \cite{hur2017survey, mehrnezhad2016nfc} or active relays \cite{wang2012implementation, korak2014power, tu2019addressing} that amplify and forward NFC signals. Active NFC-enabled devices effectively increase the hacking range to 1.5 meters \cite{LongrangeDetection2011}, with increased power consumption and hardware cost. Other work also focused on NFC eavesdropping \cite{kortvedt2009eavesdropping, wang2018information}. The  eavesdropping range of magnetic field using passive NFC was first evaluated in \cite{brown2013evaluating}, and a more powerful reader was designed to increase the communication range by 3--5 times in \cite{kirschenbaum2006build}.

\textbf{Passive Attacks in NFC:} Recent work on the design of passive attacks, primarily in the context of wireless power transfer, use the magneto-inductive (MI) principle \cite{gulbahar2017communication, dumphart2017magneto, ahmed2017magneto, vallecchi2018coupling}. There has been much work using the MI model to improve the relay systems. Theoretical optimization problems for interference zero-forcing beamforming \cite{kisseleff2014throughput, sun2012capacity} were formulated to analyze the optimal orientation and distance of relays. Binary load and frequency tuning \cite{dumphart2017magneto} show an improvement of 35.6dB in the whole passive relaying system. Lagrangian multi-user approach \cite{gulbahar2017communication} was introduced and analyzed to obtain optimal orientations maximizing received power. Other work adopted passive relays in multiple applications. MagMIMO \cite{jadidian2014magnetic} designed a multiple-input multiple-output beamforming system and passive antennas attached to remote mobile phones, achieving a charging distance of 40 centimeters. MultiSpot\cite{shi2015wireless} focused on multiple receivers, and beamformed the maximum current by inferring the number of receivers.

\textbf{Detecting Attacks:} Mechanisms of detecting the potential attacks on NFC have also been widely studied. Recent state-of-art work on detecting attacks in NFC fall into distance bounding \cite{reid2007detecting, zhuang2017energy, kilincc2017contactless}. A security framework of contactless access control \cite{kilincc2017contactless} considers the whole interaction between components, transforming the distance boundaries of access control to protect user privacy of NFC tags. Other work also used ambient sensors \cite{halevi2012secure, gurulian2017effectiveness, shepherd2017applicability} and security middleware \cite{hameed2014lightweight, parmar2017securing, shen2017new} to detect the malicious attack. Seven ambient sensors were used in \cite{gurulian2017effectiveness} to separate normal and malicious transactions that may occur at the same time. A security middleware is designed in \cite{hameed2014lightweight} to recognize CPU footprints with low latency less than a second. Further, infrared light was also adopted to prevent relay attacks in NFC \cite{gurulian2017preventing}.

Inspired by the models above, we build a practical system, \name,  that designs passive relays to break the guaranteed maximum communication distance. The total communication distance between the reader and the attacked card is enlarged to 49.6~cm using completely passive components, an improvement of 10$\times$ over commercial NFC.
% We first find the optimal separation and orientation by carrying out analysis, extensive simulations and experimentations. We develop a wearable system which can read an NFC card behind the attacker in the presence of blockages.

\textbf{Why Passive?} Our system focuses on physical attack exploiting the design of completely passive relays. First, \name\ does not consume power given that its relay coils are made from common copper wires and capacitors, making our system a low-cost and energy-efficient one. Second, our attack device can be relatively easily hidden without the power supply or the batteries, which means that the attacker can carries the attack device more stealthily when compared to the active ones. Then, though active transmitters bring us novel approaches to attack NFC, our system passively relays the query and authentication signals of the commercial reader over extended range without any active eavesdropping and data modification. Thus, it is relatively difficult to localize our relays. Furthermore, \name\ tackles the complementary problem of aiming to enlarge the communication range of NFC cards without modifying the protocol. In the absence of active transmissions, modifications to the radio signal or changes to signal timing, these make \name\ hard to detect using previous work on NFC threat mitigation.

\section{Preliminaries}\label{background}
NFC has three modes of operation: the read/write mode, card emulation mode and peer-to-peer mode. In the read/write mode, an NFC device is used as a reader, communicating with the NFC tags. This mode is frequently used in access control system or mobile payments. In the card emulation mode, an NFC device acts as a passive card and can communicate with NFC readers when it receives enough energy to wake up. NFC smartphones operate in peer-to-peer mode to share useful information, like the contact lists or pictures between  phones. In this paper, we consider NFC tags working in card emulation mode (e.g. payment cards or access cards) and  a commercial reader transmitting signals  to NFC tags.

\textbf{The NFC Protocol for Authentication:} When processing payments, an NFC reader keeps sending a query signal and waiting for the response from NFC cards. Once the NFC cards accumulate enough energy, they use load modulation to decode the reader's signal and then send a response which includes a 16-bit random number and an ID number (typically 4-7 bytes long).
%For instance, the Mifare card used in our system can be Ultralight or Classic 1k. The ID number of Ultralight has 4 bytes, while the ID number of Classic 1k has 7 bytes.
The payment process is authenticated after a successful handshake between the card and the reader. %According to different protocols for payments, different NFC cards can complete transactions with the readers.

\begin{figure}[htbp]
\begin{center}
\includegraphics[width=0.3\textwidth]{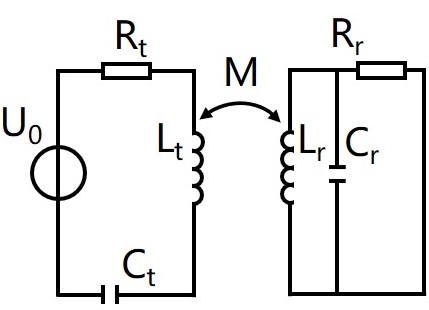}
\caption{Electromagnetic induction in NFC.}
\label{fig_magnetic}
\end{center}
\end{figure}

\textbf{Physics behind NFC:} NFC tags use electromagnetic induction to supply  power to wake up NFC tag chips. As Fig.~\ref{fig_magnetic} shows, the reader, powered by a power source $U_0$, transmits  electric and magnetic energy through the coil antenna. The coil antenna on the NFC tag then observes an induced current and magnetic field due to mutual inductance. The NFC tag has a typical LC parallel circuit.

The larger the amount of magnetic flux from the transmitter's coil that goes through the receiver's coil, the larger the induced current is at the receiver. To maximize induced current, the coils are designed to be resonant at the NFC frequencies ($13.56$ MHz) so that the inductance and capacitance: $j\omega L$ and $\frac{1}{j\omega C}$ cancel each other. We will find the optimal distance between  two coils so as to maximize the mutual inductance between them.

\begin{figure}[htbp]
\begin{center}
\includegraphics[width=0.45\textwidth]{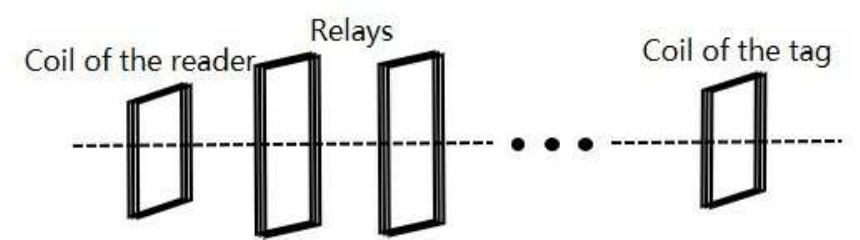}
\caption{The MCR-WPT model.}
\label{fig_mi}
\end{center}
\end{figure}

The net separation between a pair of coils for the same desired current at the receiver coil can be increased by using additional coil relays, which enhance the MCR-WPT. Passive relays help guide magnetic flux more efficiently from transmitter to receiver, allowing the receiver tag to receive the same current from a much longer distance (see Fig.~\ref{fig_mi}). This model has an equivalent circuit to mathematically calculate the mutual inductance. After obtaining the mutual inductance, we can calculate the corresponding impedance which can represent the coupling degree between two coils.

%\begin{aligned}
%Z_{i(i-1)} = \frac{\omega^2M^2}{Z + Z_{(i+1)i}}, \label{Z1}\\
%(i = 2,3,\cdots,n-1 and Z_{n(n-1)} = Z_r) \nonumber
%\end{aligned}
%where $Z = R + j\omega L+ \frac{1}{j\omega C}$
% To see how mathematically, let us abstract the passive coil relay system (Figure \ref{fig_mi}) as an equivalent model of MCR-WPT as shown in  Figure \ref{fig_coil} shows. After obtaining the mutual inductance $M$, we can calculate the corresponding impedance which can represent the coupling degree between two coils, as in Eqn. \ref{Z1} and Eqn. \ref{Z2} show. %Based on this model, we can design our passive relays by coupling and push the range limits of NFC without external power supply.

% \begin{equation}\label{Z1}
% \begin{split}
% & Z_{i(i-1)} = \frac{\omega^2M^2}{Z + Z_{(i+1)i}},\\
% (i = 2,& 3, \cdots,n-1 \ and\  Z_{n(n-1)} = Z_r),
% \end{split}
% \end{equation}
% where $Z = R + j\omega L+ \frac{1}{j\omega C}$

% \begin{equation}\label{Z2}
% \begin{split}
% & Z_{(i-1)i} = \frac{\omega^2M^2}{Z + Z_{(i-2)(i-1)}},\\
% (i & = 3, 4, \cdots,n \ and\  Z_{12} = \frac{\omega^2M^2}{Z}).
% \end{split}
% \end{equation}

The rest of this paper relies on these principles to study the effect of the human body, the coil design space (geometry, size, number of turns) as well as the circuit impedance to maximize power transfer in real-world NFC passive man-in-the-middle attacks.

%\begin{equation}\label{U1}
%\begin{split}
%& U_i = -j\omega M \frac{U_{i-1}}{Z + Z_{(i-2)(i-1)}},\\
%& (i = 2, 3, \cdots,n \ and\  U_1 = U_0),
%\end{split}
%\end{equation}
%
%Then we can obtain the received power on the tag's side as:
%\begin{equation}\label{power}
%P_r = Re \{\frac{Z_r\cdot U_{n}^2}{(Z_{(n-1)n} + Z + Z_r)^2}\}.
%\end{equation}

%\begin{figure}
%\includegraphics{magnetic.pdf}\Description{fig_magnetic}
%\caption{Electromagnetic induction in NFC.}
%\end{figure}

\section{System Model}\label{systemmodel}

In this section, we introduce the attack model, and then explain the challenges and the solutions in our system. Our design in Sec.~\ref{systemdesign} is around these key challenges.

\subsection{Attack Model}
\begin{figure}[htbp]
\begin{center}
\includegraphics[width=0.4\textwidth]{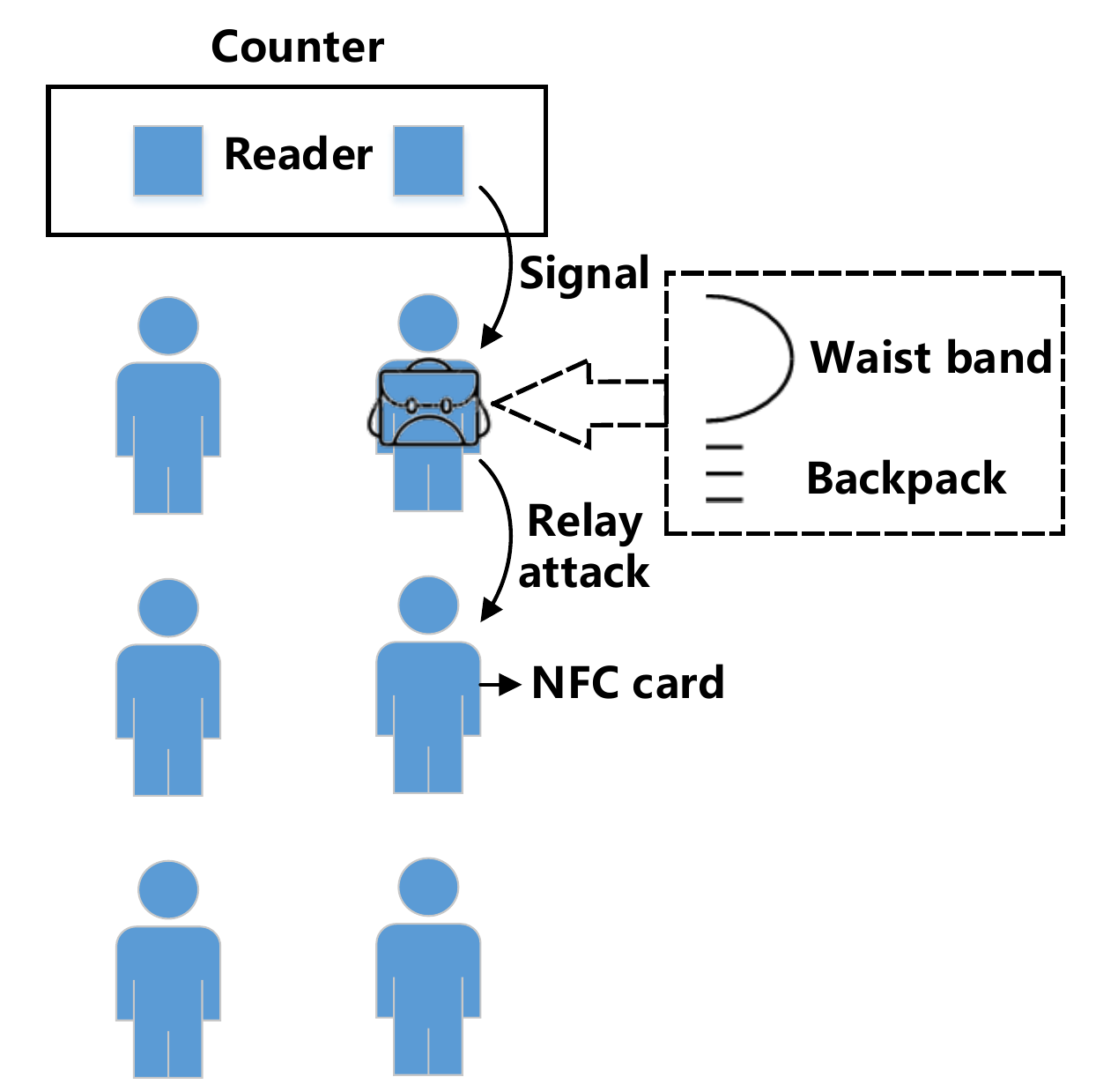}
\caption{The attack model.}
\label{attackmodel}
\end{center}
\end{figure}

As Fig.~\ref{attackmodel} shows, our attack model consists of a reader, an attacker equipped with the passive hacking device, and a victim who carries an NFC card. Our hacking device is composed of a half waist band and a backpack that contains three coils. The waist band consists of two square coils and HaBand. We use HaBand to reroute the signal pass over the surface of the human body, and use the square coils to focus the signal as much as possible. Then coils hidden in the backpack is used to convey the magnetic power to a further distance. When the attacker and the victim stand in the same line in a shop, the attacker can wear the waist band and the backpack to attack the NFC card of the person behind her.

% In this section, we introduce the attack model design, including the attack model, the coil design and how we combine the coils. In order to optimize the system and obtain the maximized communication range, we show how to design the relay coils, how to tune the capacitors on the coils, and how the distance between two coils impacts the system. We explain the challenges and the solutions in our system first.
% , and our design in the rest of the paper is around these key challenges. XXI do not get the point of the last sentenceXX

\subsection{Challenges and Solutions}

\vspace*{0.05in}\noindent\textbf{Challenge 1}: In our system, we aim to enlarge the communication range to attack the NFC cards by a wearable device. Our main goal is to transmit the signal of the reader to pass through the human body, and then to let the signal converge at the back of the attacker. However, in order to do so, we need to investigate how to determine the geometry parameters (size, number of turns, distance, orientation) that maximize the transmission range.

\vspace*{0.05in}\noindent\textbf{Solution 1}: We formulate the problem to maximize the output power in relation to the geometry parameters of the relay coils. RACA is designed to find the optimum and inform our design. Based on the algorithm, we choose a square coil geometry that senses the magnetic field generated by the signal from the reader directly and maximizes signal power relayed.

\vspace*{0.05in}\noindent\textbf{Challenge 2}:
After we design the optimal parameters to relay the energy of the NFC reader, the fast-fading signal from the reader is attenuated as it is transmitted through these coils. In particular, the signal cannot pass through the human body directly merely with the help of coupling, resulting in a limited communication range. Further, we observe that the distance that the coils can enlarge is not proportional to the size of the relay coils. This brings about the challenging problem of how to reroute the signal to circumvent the human body and focus the magnetic power towards the back of the body to the maximum extent possible.

\begin{figure*}[htbp]
\centering
\subfigure[]{
\begin{minipage}[c]{0.23\textwidth}
\centering
\includegraphics[height=3.5cm,width=4.5cm]{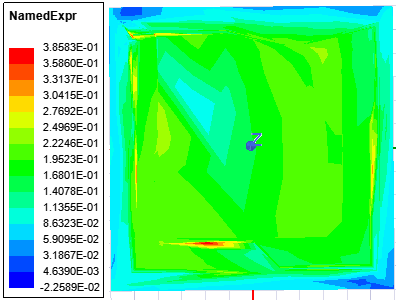}
\end{minipage}%
}
\subfigure[]{
\begin{minipage}[c]{0.23\textwidth}
\centering
\includegraphics[height=3.5cm,width=4.5cm]{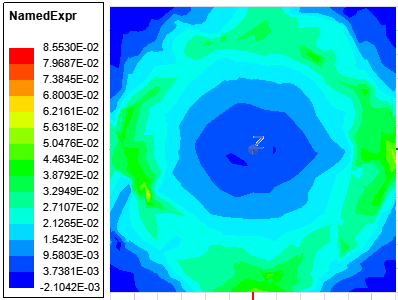}
\end{minipage}
}
\subfigure[]{
\begin{minipage}[c]{0.23\textwidth}
\centering
\includegraphics[height=3.5cm,width=4.5cm]{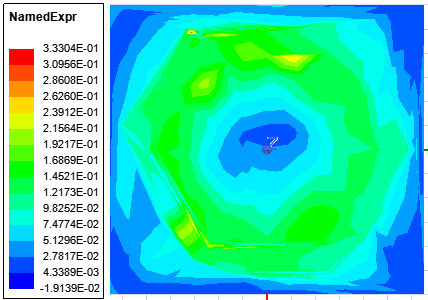}
\end{minipage}
}
\subfigure[]{
\begin{minipage}[c]{0.23\textwidth}
\centering
\includegraphics[height=3.5cm,width=4.5cm]{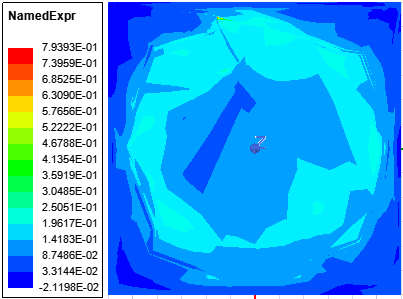}
\end{minipage}
}
\caption{The magnetic field of the: (a) square coil, (b) circular coil, (c) hexagonal coil, (d) octagonal coil.}
\label{mag_rec}
\end{figure*}

\vspace*{0.05in}\noindent\textbf{Solution 2}: To address Challenge 2, we design HaBand, a big rectangular coil which is curved to fit the waist of a human body. Since the magnetic and electric energy is relayed, the signal is transmitted across HaBand towards the back of the body. Therefore, HaBand effectively reroutes the signal and most of the power to pass over the waist of the human body, while also weakening the magnetic field as little as possible. To improve range further, we build an MCR-WPT system, where we place one of the square coils in front of the human body to connect to HaBand and place another square coil on the back of the human body.
% The four sides of the smaller coil can all generate magnetic field, making the energy gather around the area of the small coil. As a consequence, the energy which is symmetrically distributed around HaBand can be transmitted and converged as much as possible on these small coils. We combine the small coils with HaBand, and make them couple well to each other.
In this way, the system can not only transmit the signal over the surface of the human body, but also focus maximum energy to the smaller coil on the back. In order to keep enlarging the communication range in a certain direction, we build a three-square-coil box which is hidden in a backpack based on RACA.

% \textbf{Challenge 3}: When the attacker and the victim are standing in the same line, the distance between them should be more than $10cm$, which is challenging for the system to reach by passive relays. Though the smaller coil on the back can converge the greatest signal, the communication range it enlarges is limited. Therefore, the approach for the signal to continue propagating in a certain direction is essential for enlarging the whole communication range.

% \textbf{Solution 3}: In order to enlarge the communication range in a certain direction, we build a box which has three small coils and is hidden in a backpack. Considering the size of the backpack and the optimal parameter in Solution 1, the three-coil box can not only fit the thickness of the backpack, but also enlarge the communication distance more. With this backpack, the NFC card far away from the reader can be attacked.

Sec.~\ref{systemdesign} further elaborates the key challenges and solutions in the system design. Sec.~\ref{coildesign} proposes the RACA and search the optimal parameters of the coils to solve Challenge 1. Sec.~\ref{coilcombine} shows how we combine the coils and focus the power to solve Challenge 2. Sec.~\ref{evaluation} evaluates our design.

% The waist band is worn on the attacker's waist, and the box in the backpack contains three coils. In our model, we build one half band to transmit the signal from the front to the back. The band in our system includes two square coils and one big curved coil named HaBand. When the attacker and the victim are standing in the same line, the distance between the two people can be more than ten centimeters, which is difficult for NFC relays to reach.
% With the help of the backpack, the attacker can inadvertently reach the person behind her. We use HaBand to make the signal bypass the human body, and use this box hidden in the backpack to convey the magnetic power to a further distance. With the tunable capacitors on this coils, we match the optimal value of capacitor to every coil after measuring out the impedance of different coils in our system by LCR meter or vector network analyzer.

\section{System Design}\label{systemdesign}

\subsection{Coil Design}\label{coildesign}

In this section, we first determine the shape of coils based on analysis and simulations. Then we formulate the energy transmission of the attack system, which can help us determine how to design the size and geometry of the passive relays. Next, we design the RACA to obtain the optimal geometry parameters.

\subsubsection{Shape Determination}\label{ShapeDetermination}

First, we compare the magnetic field strength generated by the coil with different shapes. We assume that all the types of coils are in the uniformly changing magnetic field, and they have the same amount of magnetic flux (also the same area). We investigate square, circular, hexagonal and octagonal coils which are commonly used NFC antennas \cite{Howtodesign13.56} to observe how they affect the mutual inductance. Based on their model of mutual inductance\cite{cheng2013new,anele2015computation,tavakkoli2016analytical}, though a circular coil has the maximum magnetic flux among the four types of coil, the energy transfer is related to the power source and is always case-by-case. Thus, we simulate the four coils in HFSS to investigate the magnetic field.

In order to simulate the reader PN532 (the commercial reader we use in Sec.~\ref{implementation}) in HFSS, we build a signal source which has a rectangular coil antenna as PN532. Then, we draw a square, circular, hexagonal and octagonal coil respectively, and regulate them with the same parameters to observe the performance. The magnetic field relayed by these coils can be obtained, as Fig.~\ref{mag_rec} shows. We observe that the magnetic field relayed by the square coil is larger  than that of the others. Therefore, square relay coils are adopted in our system design.

\subsubsection{Problem Formulation}\label{Problemformulation}

\begin{figure*}[htbp]
\begin{center}
\includegraphics[width=0.8\textwidth]{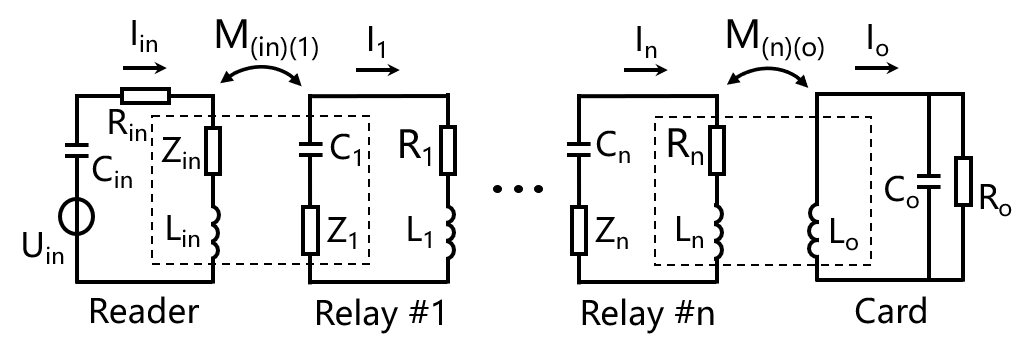}
\caption{The equivalent circuit of multiple coils in SSP MCR-WPT.}
\label{fig_coil}
\end{center}
\end{figure*}

Wireless power transfer (WPT) is recently a popular technology which is widely used in wireless charging and other applications~\cite{liu2016modeling,dai2017safe,clerckx2016waveform,dai2015survey}. Magnetically coupled inductive (MCI)-WPT and MCR-WPT are two typical techniques of WPT~\cite{hui2013critical}. MCI-WPT has limited communication range, while MCR-WPT boosts longer distance with high transmission efficiency. Therefore, MCR-WPT helps us enlarge the communication range between the reader and the card effectively. An MCR-WPT system which has a serial circuit of transmitter, several serial circuits of resonate passive relays and a parallel circuit of receiver is defined as a Serial-Serial-Parallel (SSP) MCR-WPT system. Since the NFC card has a parallel circuit, the reader-relay-card system we design is an SSP MCR-WPT system, and the equivalent circuit is shown in Fig.~\ref{fig_coil}.

In the system, we assume that we need $n$ passive relays to enlarge the communication range. $I_k (k \in \{1,2,\cdots,n\})$ represents the current of relay \#$k$, shown in Eqn.~\ref{current}. $R_k$ represents the corresponding resistance, and $Z_k$ represents the equivalent impedance. Table~\ref{notation} shows the symbols we define in \name . $\omega$ is the angular frequency and $\omega=2\pi{f}$, $f$ is the resonate frequency (13.56~MHz) of NFC, $M_{12}$ represents the mutual inductance between relay \#1 and \#2, and so on. $M_{(in)(1)}$ is the mutual inductance between the NFC reader and the relay \#1. $M_{(n)(o)}$ is the mutual inductance between relay \#n and the NFC card. The equivalent impedance due to the mutual inductance can be defined in Eqn.~\ref{impedance}, where $Z_{in}$ is the input impedance of NFC reader, $R_{o}$ is the output load of the NFC card.
% $R_{1}$, $R_{2}$, $\cdots$, $R_{n}$ represent the resistance of relay \#1, \#2, $\cdots$, \#n. $Z_{1}$, $Z_{2}$, $\cdots$, $Z_{n}$ represent the equivalent impedance $Z_{1}$, $Z_{2}$, $\cdots$, $Z_{n}$ represent the equivalent impedance.
% $I_1$, $I_2$, $\cdots$, $I_{n-1}$ represent the current of relay \#1, \#2, $\cdots$, \#n.
% $I_{in}$ is the input current on the NFC reader, $I_o$ represents the output current on the NFC card. Based on the equivalent circuit and Kirchhoff's circuit laws, we have

\begin{equation}\label{current}
\left\{\begin{aligned}
I_{1} &=\frac{\omega^{2} M_{(in)(1)} I_{in}}{R_{1}+Z_{1}} \\
I_{2} &=\frac{\omega^{2} M_{12} I_{1}}{R_{2}+Z_{2}} \\
&\cdots \\
I_{n} &=\frac{\omega^{2} M_{(n-1)(n)}I_{n-1}}{R_{n}+Z_{n}} \\ I_{o} &=\frac{M_{(n)(o)} I_{n}}{L_{o}} \end{aligned}\right.
\end{equation}

\begin{equation}\label{impedance}
\left\{\begin{aligned}
Z_{n} &=\frac{M_{(n)(o)} R_{o}}{L_{o}^{2}} \\ Z_{n-1} &=\frac{\omega^{2} M_{(n-1)(n)}^{2}}{Z_{n}+R_{n}} \\
& \cdots \\
Z_{1} &=\frac{\omega^{2} M_{12}^{2}}{Z_{2}+R_{2}} \\
Z_{in} &=\frac{\omega^{2} M_{(in)(1)}^{2}}{Z_{1}}
\end{aligned}\right.
\end{equation}

The output power $P_{o}$ is defined as

\begin{equation}
P_{o}=I_{o}^{2} R_{o}
\end{equation}
where the $I_{o}^{2}$ is the current that passes through the load. The energy transmission efficiency can be defined as

% \begin{equation}
% \begin{aligned}
% \eta=& \frac{P_{o}}{I_{in}^{2}\left(Z_{p s}+R_{p}\right)} \cdot \\ & \frac{\omega_{0}^{4} M_{p s}^{2} M_{s r}^{2} M_{r l}^{2} L_{l}^{2} R_{o}}{h_{1}\left(\omega_{0}^{2} M_{p s}^{2} L_{l}^{2} R_{r}+\omega_{0}^{2} M_{p s}^{2} M_{r l}^{2} R_{o}+h_{1} R_{p}\right)} \end{aligned}
% \end{equation}

\begin{equation}
\begin{aligned}
\eta=& \frac{P_{o}}{P_{in}}
\end{aligned}
\end{equation}
where $P_{in} = (Z_{in}+L_{in})I_{in}^2$.

Two coils that are put together is always with lateral, axial and angular misalignment\cite{anele2015computation}. The model of mutual inductance is $M=\frac{2 N_{1} N_{2} \mu_{0}}{\pi} \sqrt{S_{1} S_{2}} \int_{0}^{\pi} \frac{\left[\cos \theta-\frac{d}{S_{2}} \cos \phi\right] \Psi(k)}{k \sqrt{V^{3}}} d \phi$, where $S_{1}$ and $S_{2}$ represent half the side length of two adjacent square coils, $N_{1}$ and $N_{2}$ represent the number of turns of the two coils. $\alpha=\frac{S_{2}}{S_{1}}$, $\beta=\frac{c}{S_{1}}$, $\xi=\beta-\alpha \cos \phi \sin \theta$, $k^{2}=\frac{4 \alpha V}{(1+\alpha V)^{2}+\xi^{2}}$, $\Psi(k)=\left(1-\frac{k^{2}}{2}\right) K(k)-E(k)$, $V=\sqrt{1-\cos ^{2} \phi \sin ^{2} \theta-2 \frac{d}{R_{S}} \cos \phi \cos \theta+\frac{d^{2}}{S_{2}^{2}}}$, $K(k)=\int_{0}^{\frac{\pi}{2}} \frac{1}{\sqrt{1-k^{2} \sin ^{2} \theta}} d \theta$, $E(k)=\int_{0}^{\frac{\pi}{2}} \sqrt{1-k^{2} \sin ^{2} \theta} d \theta$. We observe that the model of mutual inductance depends on the lateral misalignment $d$, the axial misalignment $c$, and the angular misalignment $\theta$.
% The mutual inductance between two coils in our model should be shown as Eqn.~\ref{eqn_dbl_number}.
Next we use $S_k (k \in \{1,2,\cdots,n\})$ to represent the side length of relay \#$k$. $N_{k}$ is the corresponding number of turns. $S_{in}$ is the side length of the NFC reader, and $S_{o}$ is the side length of the NFC card. $N_{in}$ is the number of turns of the NFC reader, while $N_{o}$ is the number of turns of the NFC card. We define $d_0, \cdots, d_n$ respectively to represent the lateral misalignment between any adjacent two coils of the $n+2$ coils (NFC reader, NFC card, and $n$ relays). Besides, $c_0, \cdots, c_n$ and $\theta_0, \cdots, \theta_n$ also represent the corresponding axial and angular misalignment. We also define all the above symbols in Table~\ref{notation}.

Based on the models above, in our problem, we maximize the output power of this coil system in Eqn.~\ref{max} below:

\begin{equation}\label{max}
\max \quad P_{o}
\end{equation}

\begin{equation}\label{constraints}
\text { s.t. }\left\{\begin{array}{l}{\eta^n \geqslant \eta_{0}} \\ {I_{in} < I_{0}} \\ {S_k \leqslant S_{0}} \end{array}\right.
\end{equation}

According to Eqn.~\ref{constraints}, in order to guarantee the signal transmission, the transmission efficiency $\eta^n$ when we use $n$ relays should be equal to or larger than the threshold $\eta_0$. Besides, due to the power constraint of FCC, the current of the NFC reader should be less than the threshold $I_0$. Further, due to the area constraint of human body, $S_k$ should be equal to or less than the threshold $S_0$.

This optimization problem is formulated as a nonlinear programming, which is non-convex and includes extensive nonlinear constraints. We cannot map our problem to an underlying convex problem and adopt convex optimization methods directly. Conducting the nonlinear programming solvers will bring about largely local optimum. Considering that parameter settings under experiments could be step by step, we try to analyze feasible regions and narrow them down as much as possible. Then we design RACA which will be introduced as follows to solve this problem.

\begin{table}[!hpb]
\caption{Notations}
\label{notation}
  \centering
  \begin{tabular}
    {@{}cc@{}} \toprule \midrule
    %\multicolumn{2}{c}{Item} \\ \cmidrule(r){1-2}
    Symbol & Definition \\
    \midrule
    \midrule
    $n$ & The number of relays \\
    \midrule
    $k$ &  Relay \#$k$ ($k \in \{1,2,\cdots,n\}$) \\
    \midrule
    $I_k$ & The current of relay \#$k$ \\
    \midrule
    $I_{in}$ & The input current on the NFC reader \\
    \midrule
    $I_o$ & The output current on the NFC card \\
    \midrule
    $R_k$ & The resistance of relay \#$k$ \\
    \midrule
    $R_o$ & The output load of the NFC card \\
    \midrule
    $Z_k$ & The equivalent impedance of relay \#$k$ \\
    \midrule
    $Z_{in}$ & The input impedance of NFC reader \\
    \midrule
    $L_{in}$ & The inductance of the NFC reader \\
    \midrule
    $L_{o}$ & The inductance of the NFC card \\
    \midrule
    $N_{in}$ & The number of turns of the NFC reader \\
    \midrule
    $N_{k}$ & The number of turns of relay \#$k$ \\
    \midrule
    $N_{o}$ & The number of turns of the NFC card \\
    \midrule
    $S_k$ & The side length of relay \#$k$ \\
    \midrule
    $S_{in}$ & The side length of the NFC reader \\
    \midrule
    $S_{o}$ & The side length of the NFC card \\
    \midrule
    $f$ & The resonate frequency  \\
    \midrule
    $\omega$ & The angular frequency $\omega=2\pi{f}$  \\
    \midrule
    $M_{12}$ & The mutual inductance between relay \#1 and \#2  \\
    \midrule
    $M_{(in)(1)}$ & \tabincell{c}{The mutual inductance between\\the NFC reader and relay \#1}    \\
    \midrule
    $M_{(n)(o)}$ & \tabincell{c}{The mutual inductance between \\relay \#n and the NFC card}  \\
    \midrule
    $P_{in}$ & The input power  \\
    \midrule
    $P_o$ & The output power  \\
    \midrule
    $d_0, \cdots, d_n$ & The lateral misalignment between two adjacent  coils  \\
    \midrule
    $c_0, \cdots, c_n$ & The axial misalignment between two adjacent  coils  \\
    \midrule
    $\theta_0, \cdots, \theta_n$ & The angular misalignment between two adjacent coils  \\
    \midrule
    $\eta^n$ & The transmission efficiency when we use $n$ relays  \\
    \midrule
 \bottomrule
  \end{tabular}
\end{table}

\subsubsection{\name-Ant Colony Algorithm}\label{Parametersearch}

Ant Colony Algorithm (ACA) is a heuristic algorithm that is inspired by the pheromone trail laying and following behavior of some ant species~\cite{dorigo2019ant}. Artificial ants find the best path by sensing the density of the pheromone left by the ant colony. Inspired by this, we design RACA to obtain the optimal parameters that maximize the output power. We first divide the feasible areas into several scatter sets, narrowing down the feasible areas by taking account for real experiments. For example, we do not have to fix the iteration step 1~mm, since 1~mm is negligible in real life. In order to reduce the computational complexity, $n$ should increase step by step, so we treat it as an initial parameter. In RACA, we initiate the corresponding geometric parameters, the number of iterations $T$, the volatilization coefficients $\rho$, the transition probability set $Q_0$ and the searching area set $\mathcal{A}$. Then, the randomly chosen parameters are input to calculate the initial \name\ pheromone $\tau$ which describes the contribution to maximize the output power. The parameters with larger output power are given larger transition probability. We calculate the output power and update the parameters based on transition probability every loop. Algorithm \ref{alg:alg1} shows the details. $P_o^{n*}$ is defined as the maximum output power when we use $n$ relays ($n=1,2,3,\cdots$), and we use $\eta^n$ to represent the transmission efficiency when we use $n$ relays. The local optimum can be avoided through $T$ iterations.

%   \item[\textbf{Input:} $S_{1}^0, \cdots, S_{n}^0, N_{1}^0, \cdots, N_{n}^0, d_{1}^0, \cdots, d_{{(n+1)}}^0, $]
%   \item[ \quad \quad \quad $c_{1}^0, \cdots, c_{{(n+1)}}^0, \theta_{1}^0, \cdots, \theta_{{(n+1)}}^0$ ]
% \item[\textbf{Output:} $S_{1}^*, \cdots, S_{n}^*, N_{1}^*, \cdots, N_{n}^*, d_{1}^*, \cdots, d_{{(n+1)}}^*,$]
% \item[ \quad \quad \quad \  $c_{1_0}^*, \cdots, c_{{(n+1)}}^*, \theta_{1}^*, \cdots, \theta_{{(n+1)}}^*$ ]

\begin{algorithm}
  \caption{\name-Ant Colony Algorithm}\label{alg:alg1}
  \begin{algorithmic}[1]
  \item[\textbf{Initiate} $P_o^{n*} = 0, n=1,2,3,\cdots, T, \rho, Q_0, \mathcal{A}$]
  \item[ \quad \quad \quad \quad \quad \quad \quad \quad \quad \quad \quad \quad \quad \quad \Comment{Initialize the parameters}]
  \item[\textbf{Input:} $S_k, N_k, d_0, \cdots, d_n, c_0, \cdots, c_n, \theta_0, \cdots, \theta_n$]
  \item[ \quad \quad \quad \quad \quad \quad \quad \quad \quad \quad  \Comment{Input parameters in feasible areas}]
  \item[\textbf{Output:} $S_k^*, N_k, d_0^*, \cdots, d_n^*, c_0^*, \cdots, c_n^*, \theta_0^*, \cdots, \theta_n^*$ ]
  \item[\quad \quad \quad  \Comment{Optimal parameters to maximize the output power}]
  \item[\textbf{Loop:}]
  \State $\tau = CALfun(\rho, Q_0, \mathcal{A})$
  \item[\quad \quad \quad \quad \quad \quad \quad \quad \quad \quad \quad \Comment{Calculate the initial pheromone}]
  \State \textbf{while} $\eta^{n} \geqslant \eta_{0}, I_{in} < I_{0}, {S_k \leqslant S_{0}}, iteration\ times \leqslant T$ \quad \textbf{do}
  \State \quad Calculate $P_o^{n}$
  \State \quad \textbf{if} $P_o^{n} > P_o^{n*}$
  \State \quad \quad $P_o^{n*} = P_o^{n}$
  \item[\quad \quad \quad \quad \quad \quad \quad \quad \quad \Comment{Search the maximum output power}]
  \State \quad  \textbf{end if}
  \State \quad Update $S_k, N_k, d_0, \cdots, d_n, c_0, \cdots, c_n, \theta_0, \cdots, \theta_n$
  \State \quad Update $\tau = (1 - \rho) \times \tau + CALfun(\rho, Q_0, \mathcal{A})$
  \State \textbf{end while}
  \State $ S_k^* \gets S_k$
  \State $ N_k^* \gets N_k$
  \State $ d_0^*, \cdots, d_n^* \gets d_0, \cdots, d_n $
  \State $ c_0^*, \cdots, c_n^* \gets c_0, \cdots, c_n $
  \State $ \theta_0^*, \cdots, \theta_n^* \gets \theta_0, \cdots, \theta_n $
    \end{algorithmic}
\end{algorithm}

We observe that the value of mutual inductance keeps dropping when the angular misalignment increases, thus when the two coils are parallel, the largest mutual inductance is obtained. Moreover, when the lateral misalignment increases, the mutual inductance between the two coils drops first and gradually increases to almost 0. With the axial misalignment increases, the mutual inductance reduces and approaches to 0.

We then tune the side length, the number of turns, the lateral, axial, and angular misalignment to study the output power. With the side length, the number of turns, or the axial misalignment of the square coils increasing, the output power increases first and then drops gradually. Further, the output power is inversely proportional to the lateral and angular misalignment. When the lateral misalignment is equal to 0 and the axial misalignment is equal to half the side length of the square coil, we have the largest output power. As a consequence, we build the square coils 10~cm$\times$10~cm with three turns and a separation of two-millimeter between every two copper wires.

\begin{figure}
\centering
\subfigure[]{
\begin{minipage}[c]{0.25\textwidth}
\centering
\includegraphics[height=3.5cm,width=4.8cm]{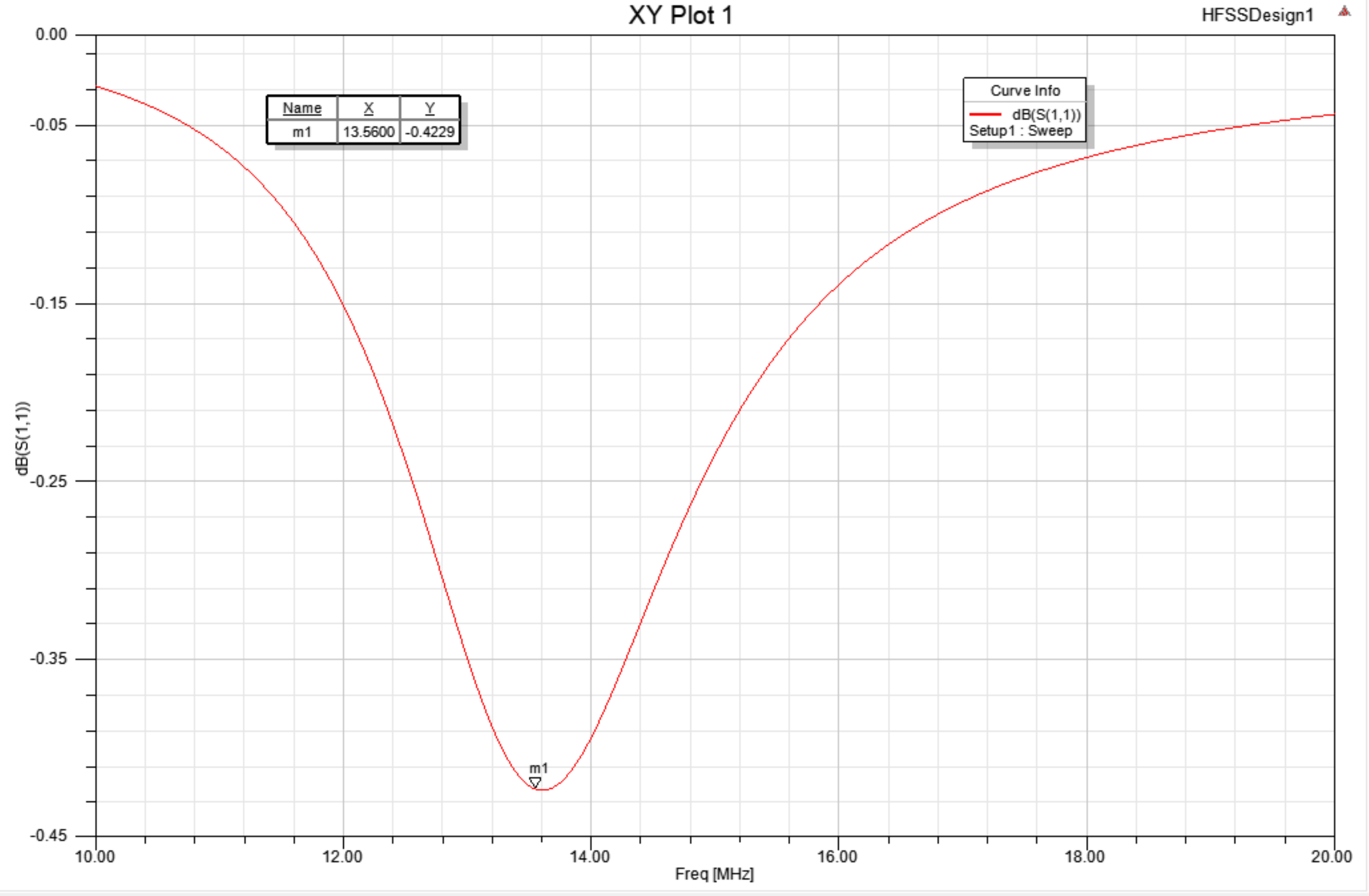}
\end{minipage}%
}
\subfigure[]{
\begin{minipage}[c]{0.21\textwidth}
\centering
\includegraphics[height=3.5cm,width=3.5cm]{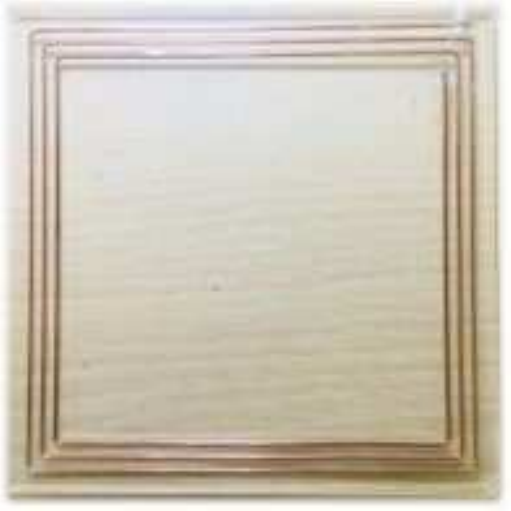}
\end{minipage}
}
\caption{(a) The return loss S11 under 51pF. (b) The prototype of the square coil.}
\label{fig_optim}
\end{figure}

After we obtain the optimal parameters, we then build the model in HFSS and regulate the capacitor.
%Changing the value of the capacitor, we can see the magnetic changing in Figure \ref{fig_capa}.
We change the capacitor's capacitance progressively from 10~pF to 90~pF. Initially, With the capacitance increasing, the return loss increases  and has a peak at 51~pF. Then it gradually drops when the capacitance keeps increasing. Such a result validates that the coil can transmit the largest magnetic power at the resonant frequency 13.56~MHz, as Fig.~\ref{fig_optim}(a) shows.

% The magnetic field under $51~pF$ can be seen in Figure \ref{fig_optim}(a) and the return loss parameter S11 can be seen in Figure \ref{fig_optim}(b). Figure \ref{fig_optim}(b) shows that the resonant frequency of this square coil is approximate to $13.56~MHz$, which is consistent with the frequency of NFC signal. Due to the ideal environment in the simulation model, we can see the difference between the simulation result and the analyzed result based on Equation \ref{eq:frequency}.

\begin{equation}\label{eq:frequency}
f = \frac{1}{2 \pi \sqrt{LC}}.
\end{equation}

We design our prototype by using copper wires and capacitors, shown as Fig.~\ref{fig_optim}(b). Based on Eqn.~\ref{eq:frequency} and extensive experiments, we match a capacitance of 60~pF for the system to achieve the largest distance expanded. Taking into account that the commercial capacitors have a 10\% error, the difference of results among the analysis, the simulation and the real coils can be acceptable.

\subsection{Coil Combination}\label{coilcombine}

\subsubsection{HaBand Design}\label{habanddesign}

Since the signal cannot go through the human body directly, we build HaBand which has a dimension of 10~cm$\times$30~cm. It also has three turns and a two-millimeter separation between every two copper wires. We match the appropriate capacitance to it based on Eqn.~\ref{eq:frequency}. Then, we change its curvature to match the curvature of human waist.

After we manually build the 10~cm$\times$10~cm square coils and the 10~cm$\times$30~cm HaBand, we combine them and build an MCR-WPT system,  focusing the signal at the back of the human body as much as possible.
% From Sec.~\ref{coildesign} we observe that the transmission efficiency is related to distance between the resonant coils. The model and results of transmission efficiency are also applicable to the coils in the same plane (the axial and angular misalignment are equal to zero).
When two coils horizontally laying in the same plane couple each other too close, one of them is the source of interference to the other, weakening the whole magnetic fields. When we gradually separate the two coils in the same plane, the coupled magnetic field would increase first and then drop. The parameters that describe the combined magnitude of the magnectic field in HFSS also validate this observation. We also evaluate this result by using Universal Software Radio Peripheral (USRP) in Sec.~\ref{evaluation}. Indeed, this is the reason why we need a small square coil to absorb the energy from the NFC reader, before we use HaBand and another small square coil. Together, our approach focuses the signal and magnetic power in the desired direction around the attacker's body.

% \begin{figure}[htbp]
% \begin{center}
% \includegraphics[width=0.15\textwidth]{figure/coil_prototype}
% \caption{Our prototype of the square coil.}
% \label{coil_prototype}
% \end{center}
% \end{figure}

% In order to couple well to the rectangular coils, HaBand has the same turns and separation of copper wires as the rectangular ones. Here we build our system a MR-WPT system, which combine the square coils and HaBand.

% We then build a model in HFSS to validate it. Actually, when the side of the first coil perfectly overlaps with one side of the other coil, we obtain the best magnetic power. We start from letting the two coils overlap with each other, then we slowly change the distance and observe the parameters S11, S21 and the NearETotal. S11 is return loss which describes the energy transmitted by port 1 and reflected to port 1, while S21 is transmission loss which describes the energy transmitted from port 1 to port 2. The maximum NearETotal is the combined magnitude of the electric field components. When we set the distance parameter 96~mm (the two sides perfectly overlap with each other), the system generates a larger magnetic field. We will evaluate this result by using universal software radio peripheral (USRP) in Sec. \ref{evaluation}.

In order to obtain the whole optimization in our system, it is important to make sure that every coil has the best performance. Considering that there are some errors among different coils and different capacitors, we use the tunable capacitors which can be effortlessly hand-tuned to be matched to every unique coil.

\subsubsection{Backpack design}\label{backpack}

After converging the signal at the back of the human body and regulating the optimal value of capacitors, we want to find the best way to keep the signal transmitting farther. Thus more coils need to be added to this system. As we know from Sec.~\ref{coildesign}, when we gradually separate two coils, the magnetic power increases first and reaches a peak, then it finally drops. We add three more coils based on the optimal distance between two coils we obtain. We also test the performance by changing the distance in HFSS. The results show that the two-coil system has a better performance when the distance is between 5~cm and 6~cm, which validates the analysis in Sec.~\ref{coildesign}. In Sec.~\ref{evaluation}, we introduce how we evaluate this range improvement using a software radio.

\begin{figure}[htbp]
\begin{center}
\includegraphics[width=0.25\textwidth]{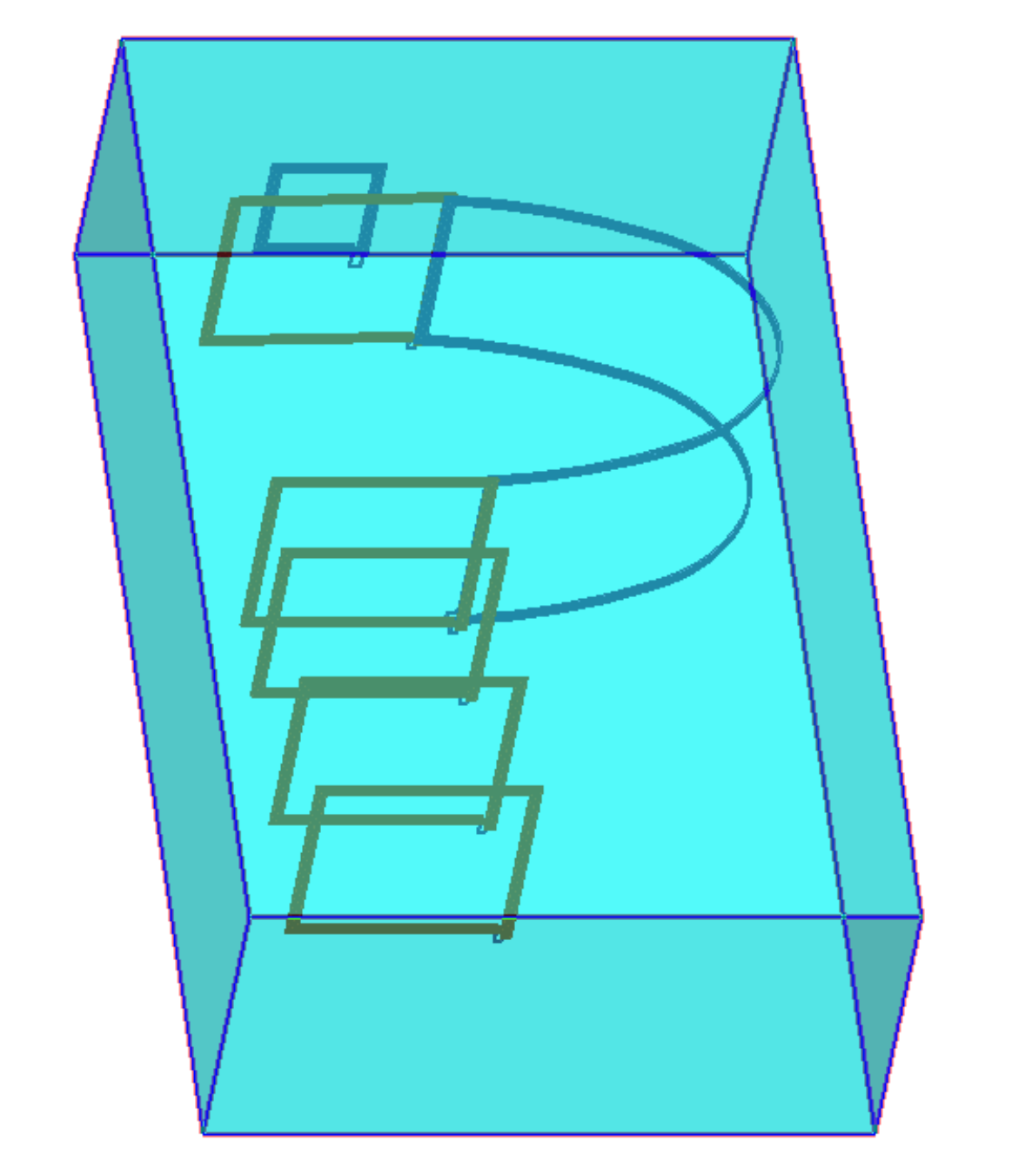}
\caption{The \name\ simulation model in HFSS.}
\label{circle}
\end{center}
\end{figure}

Orientation can also impacts the generated magnetic field. In order to verify the optimal orientation of the coils that we put in the backpack, we change the orientation of the coils in HFSS model. The system is expected to get the optimal performance when the two coils totally face each other. This result is consistent with that in Sec.~\ref{coildesign}. We let the distance between the two coils be 5~cm and change the orientation parameter of the second coil from 0~deg to 180~deg. When the orientation parameter is 0~deg, the system can generate the largest magnetic field. In Sec.~\ref{evaluation}, we also use USRP to evaluate our system performance by changing the orientation of coils. After we obtain the optimal distance between the coils and the orientation of the coils, a backpack is used to hide the additional three coils. As a consequence, more distance enlargement to hack people's cards can be achieved.

\begin{figure}[htbp]
\begin{center}
\includegraphics[width=0.48\textwidth]{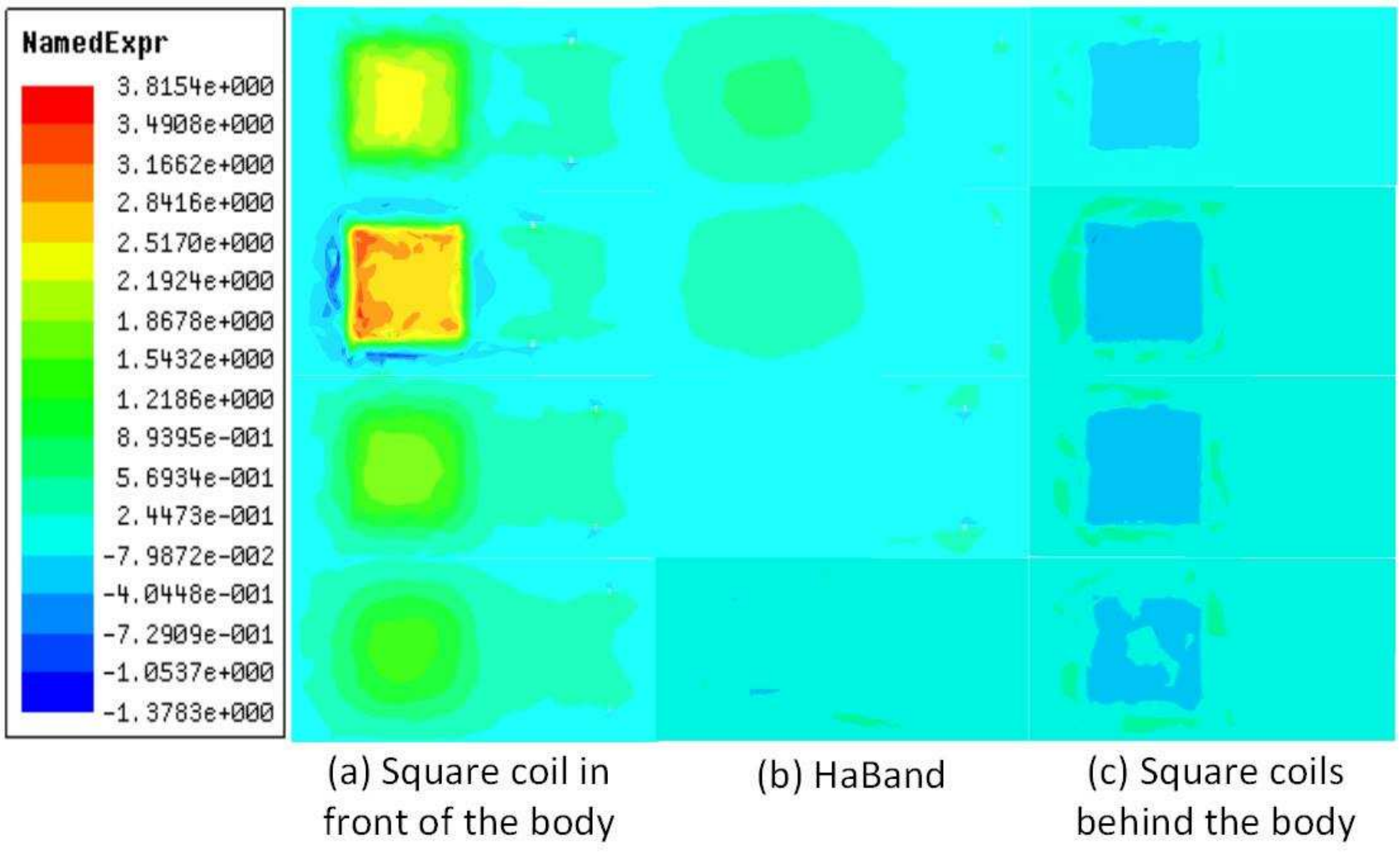}
\caption{The cross-sectional planes of the magnetic field in \name\  system.}
\label{plane}
\end{center}
\end{figure}

Based on the results above, we build our attack system in HFSS to observe the transmission process of the magnetic field. As Fig.~\ref{circle} shows, the small rectangle in front of the whole model acts as the commercial reader which generates magnetic signals. Then, we introduce our system which consists of the square coils and HaBand. Another three square coils hidden in the backpack keep delivering the signal and energy far away from the band. We introduce several cross-sectional planes across the model to see the transmission process of the magnetic power. Column(a) in Fig.~\ref{plane} illustrates the magnetic field around the square coil beside the reader. Column(b) introduces that the magnetic field generated by the square coil is gradually weaker and largely transmitted on HaBand. From Column(c), the magnetic field begins to converge on the square coils behind HaBand. In summary, the magnetic field is first absorbed by the square coil in front of the human body, and then it is transmitted by HaBand. It finally converges towards the coils in the backpack. These results validate our analysis from before.

\section{Implementation}\label{implementation}

% In this section, we implement the proposed system.

% \subsection{\name\ Prototype}

We build the testbed that consists of a reader, an attacker equipped with \name , and a victim who carries an NFC card. Our testbed is built in a 4.5~m$\times$4.5~m office area, shown as Fig.~\ref{fig_implement}. We place the PN532 NFC reader which is connected to a personal computer (PC) through the serial port on the desk. PN532 NFC shield is attached to Arduino through $\text{I}^2\text{C}$ interface. One person acts as the attacker, and the other who holds NFC cards acts as the victim. The waist band is worn on the attacker's waist and the three-coil box is hidden in a backpack. Through the serial port software, the PC on the desk reads the information about the hacked card, including its unique ID and whether it can be authenticated or not.

% \begin{figure}[htbp]
% \begin{center}
% \includegraphics[width=0.4\textwidth]{figure/implementation1}
% \caption{The \name\ testbed.}
% \label{implementationn}
% \end{center}
% \end{figure}

\begin{figure}
\centering
\subfigure[]{
\begin{minipage}[c]{0.23\textwidth}
\centering
\includegraphics[height=4cm,width=4cm]{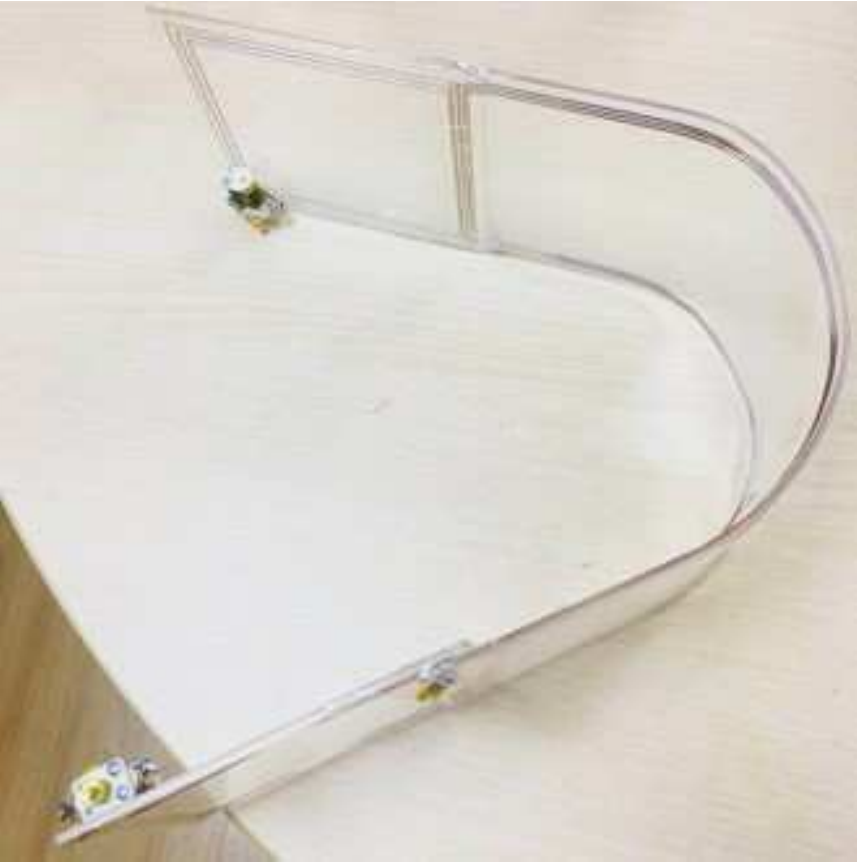}
\end{minipage}%
}
\subfigure[]{
\begin{minipage}[c]{0.23\textwidth}
\centering
\includegraphics[height=4cm,width=4cm]{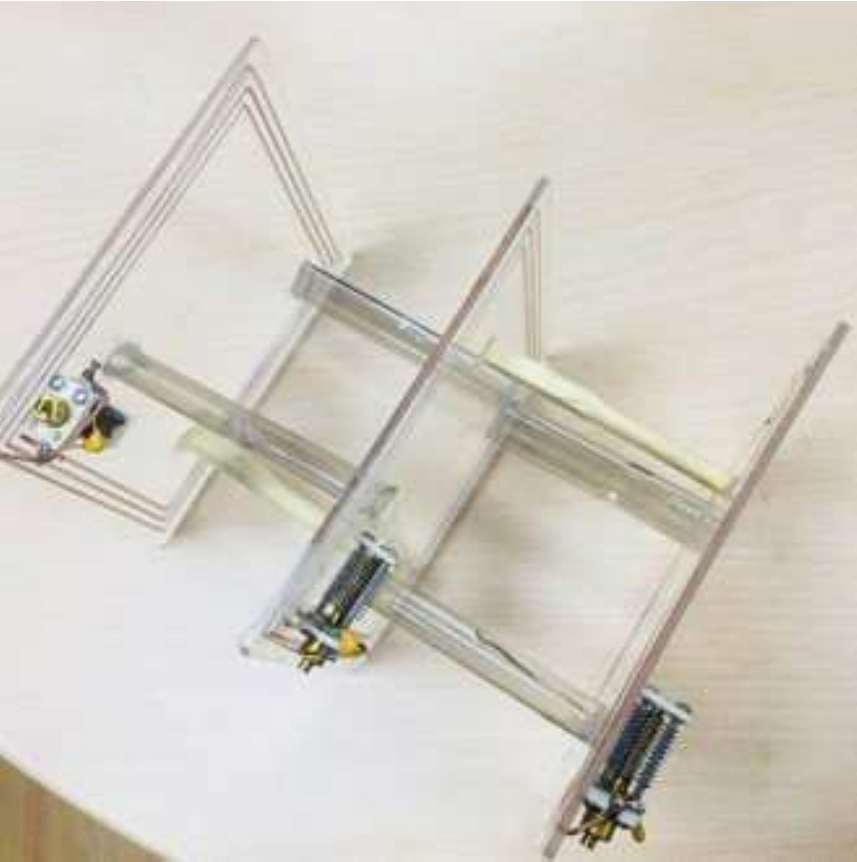}
\end{minipage}
}
\caption{(a) The waist band. (b) The three coils put into the backpack.}
\label{waistband}
\end{figure}

Fig.~\ref{waistband} shows the design of the attack device. It is composed of the waist band and the three-coil box placed into the backpack. All the coils are fixed on acrylic plates which do not introduce interference. HaBand has a dimension of 10~cm$\times$30~cm, while the square coils have a dimension of 10~cm$\times$10~cm. We make HaBand curved by hot-melt technique to fit around the human body's waist. In order to guarantee the largest magnetic field, we regulate every tunable capacitors to achieve the best performance of the device. Fig.~\ref{waistband}(b) also shows our design on the box we put into the the backpack. From the results in Sec.~\ref{systemdesign}, we first set the distance of every two of the square coils 5~cm. Then, we make every two of them totally face each other, i.e., we set the orientation 0~deg.

% \subsection{Testbed Experiments}

In our system, we conduct five types of experiments.
(1) Experiments of two-coil system. We build two square coils by using copper wires and capacitors. In this two-coil system, we deploy USRP to catch the received power. The distance between the two coils, the distance between the NFC tag and the last coil, and the orientation of the coils are changed during the process. We compare the results and optimize the parameters. (2) Experiments of the number of coils. We arrange the centroid of the square coils be in a straight line (like dominoes), and investigate the farthest distance that our system can enlarge by simply adding more square coils to our system. (3) Impacts of human body. We deploy our whole testbed on a human body to find the impact of the human body on the received power. We then deploy the Ferrite board on the human body to investigate the impacts on the whole distance. (4) Experiments amidst multipath. We evaluate the impacts of clothing and the impacts of multiple cards on the performance. Besides, considering the interference from the moving people, we add interference from iron plates, mobile phones and keys to our system to test the robustness of \name . (5) Experiments on commercial smartphones. We adopt commercial NFC-based smartphones as the commercial readers to test the feasibility of our system.

% When the band is worn on the human body, the distance boosted by the rectangular coil on the back can be $11.5cm$. With this box, the distance between the band and the people being hacked can reach $26cm$.
% With the attack device worn on the human body, the whole distance between the reader and the attacked card can be 48.5~cm, which is as much as ten times the communication distance of the existing NFC system.

\section{Evaluation}\label{evaluation}

In this section, we use USRP to validate the results we obtain in Sec.~\ref{systemdesign} and Sec.~\ref{implementation}. We use the USRP with the DLP radio frequency identification (RFID) antenna to listen to the real signal from the reader PN532. The DLP RFID antenna also works at a frequency of 13.56MHz. This USRP system provides us the received normalized voltage to describe the energy transfer. We first use these devices to listen to the signal generated by the reader and the response from the NFC card. We find that when the normalized voltage is larger than 0.22, the card can always decode the query signal from the reader and send the response.

\subsection{Two-coil System}\label{two-coil-system}

\begin{figure}[htbp]
\begin{center}
\includegraphics[width=0.45\textwidth]{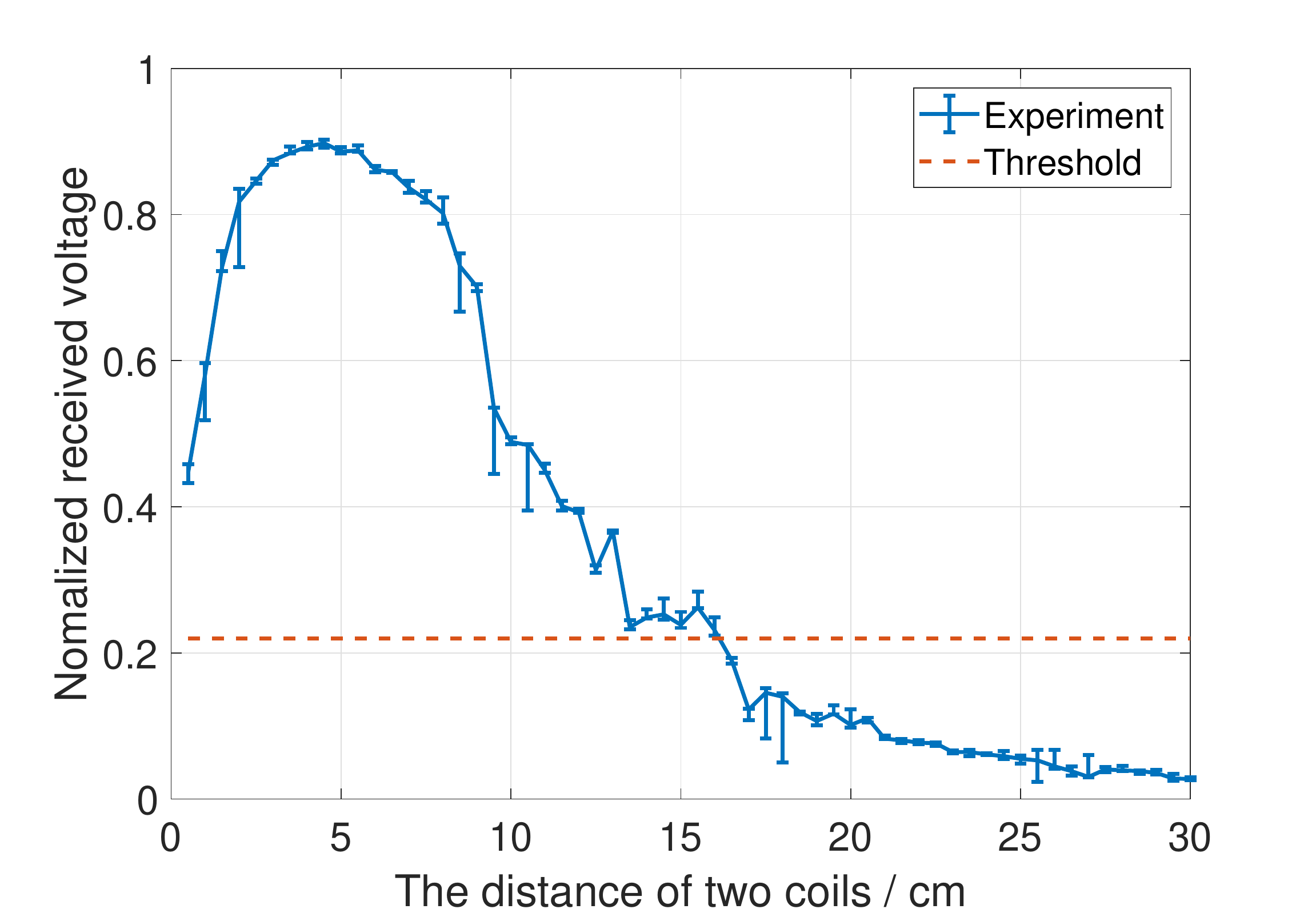}
\caption{The received voltage vs. the distance between two coils.}
\label{separation}
\end{center}
\end{figure}

\begin{figure}[htbp]
\begin{center}
\includegraphics[width=0.45\textwidth]{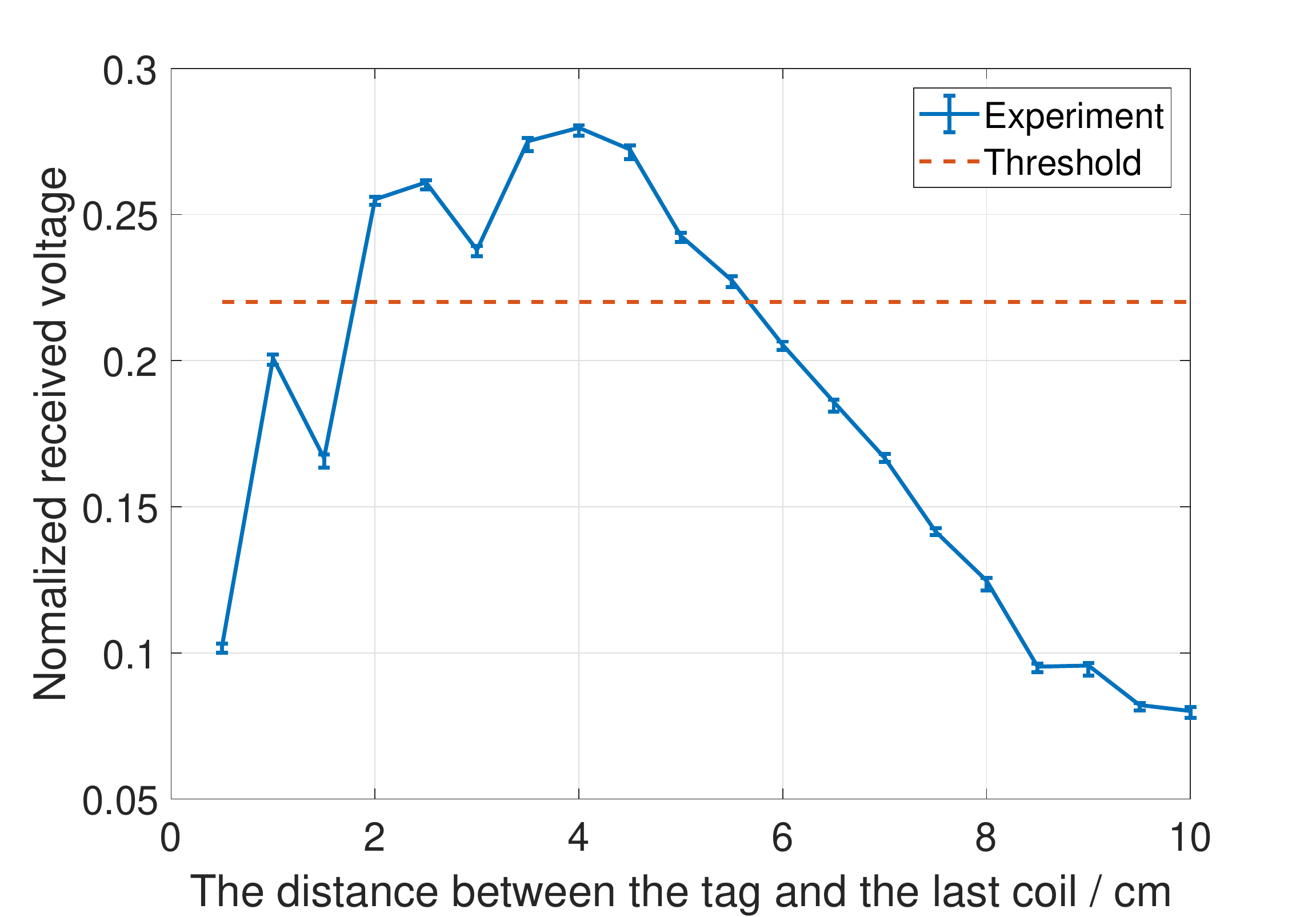}
\caption{The received voltage vs. the distance between the tag and the last coil.}
\label{card}
\end{center}
\end{figure}

We let two coils totally face each other and change the distance between them. We bind the DLP RFID antenna and the second coil together, minimizing the distance between the DLP RFID antenna and the second coil. From the results in Fig.~\ref{separation}, when the distance between the two coils is around 5~cm (half the side length of the coils), the normalized voltage received by the DLP RFID antenna reaches the peak, which means that the second coil can relay the largest power when the distance is 5~cm. The result is consistent with that of the analysis and simulation. We observe that the farthest communication distance is 16~cm in Fig.~\ref{separation}. Further, we fix the distance between the two coils 16~cm, and use the DLP RFID antenna as an NFC card. We slowly move the DLP RFID antenna far away from the second coil. As Fig.~\ref{card} shows, when the card moves, the received power reaches a peak at the distance of 4~cm, then it begins to decrease. That is to say, the second coil cuts the maximum magnetic lines and generates the maximum magnetic field at 4~cm. We observe that the error of normalized voltage in Fig.~\ref{separation} is larger than that Fig.~\ref{card}. This stems from the reason that the DLP RFID antenna is being too close to the second relay coil in Fig.~\ref{separation} and brings more interference to our system. When we fix the distance between the two coils 16~cm, we add the third relay coil to this two-coil system and find that the NFC card cannot respond to the reader anymore. This verifies our analysis that the distance between two adjacent coils should be fixed to maximize the power transfer.

% \begin{figure*}
% \centering
% \subfigure[]{
% \begin{minipage}[c]{0.23\textwidth}
% \centering
% \includegraphics[height=3.5cm,width=4.7cm]{figure/separation1.eps}
% \end{minipage}%
% }
% \subfigure[]{
% \begin{minipage}[c]{0.23\textwidth}
% \centering
% \includegraphics[height=3.5cm,width=4.6cm]{figure/card1.eps}
% \end{minipage}
% }
% \subfigure[]{
% \begin{minipage}[c]{0.23\textwidth}
% \centering
% \includegraphics[height=3.5cm,width=4.7cm]{figure/overlay1.eps}
% \end{minipage}%
% }
% \subfigure[]{
% \begin{minipage}[c]{0.23\textwidth}
% \centering
% \includegraphics[height=3.5cm,width=4.7cm]{figure/orientation1.eps}
% \end{minipage}
% }
% \caption{The received voltage due to different: (a) separation of two coils, (b) distance between the tag and the last coil. The received voltage due to: (c) the distance between the two sides of the coils, (d) the orientation of the second coil.}
% \label{usrp_card2}
% \end{figure*}

\begin{figure}[htbp]
\begin{center}
\includegraphics[width=0.45\textwidth]{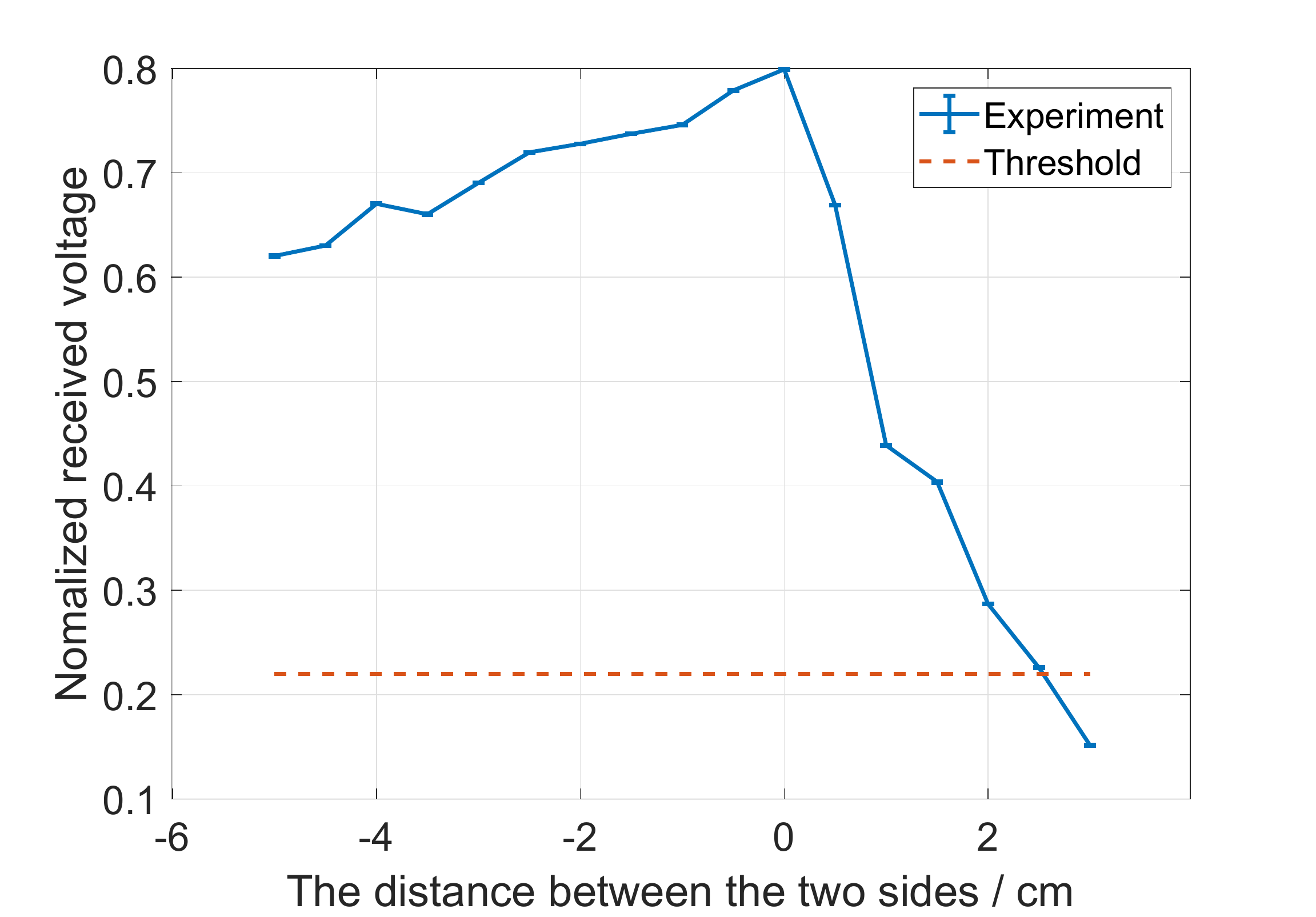}
\caption{The received voltage vs. the distance between the two sides of the coils.}
\label{overlay}
\end{center}
\end{figure}

We then place two coils horizontally in the same plane, and we fix the distance between the card and the coils to 4~cm. We progressively move one of the coils horizontally in the plane and observe the received normalized voltage. Shown as Fig.~\ref{overlay}, when the 10~cm side of the two coils are precisely next to each other, we define the distance as 0~cm. The distance is positive when the two coils become apart, and the distance is negative when they overlap. When the distance between the two sides of the coils is 0~cm, the system has the best performance. This is the reason why we compactly attach the square coils and HaBand.

\begin{figure}[htbp]
\begin{center}
\includegraphics[width=0.45\textwidth]{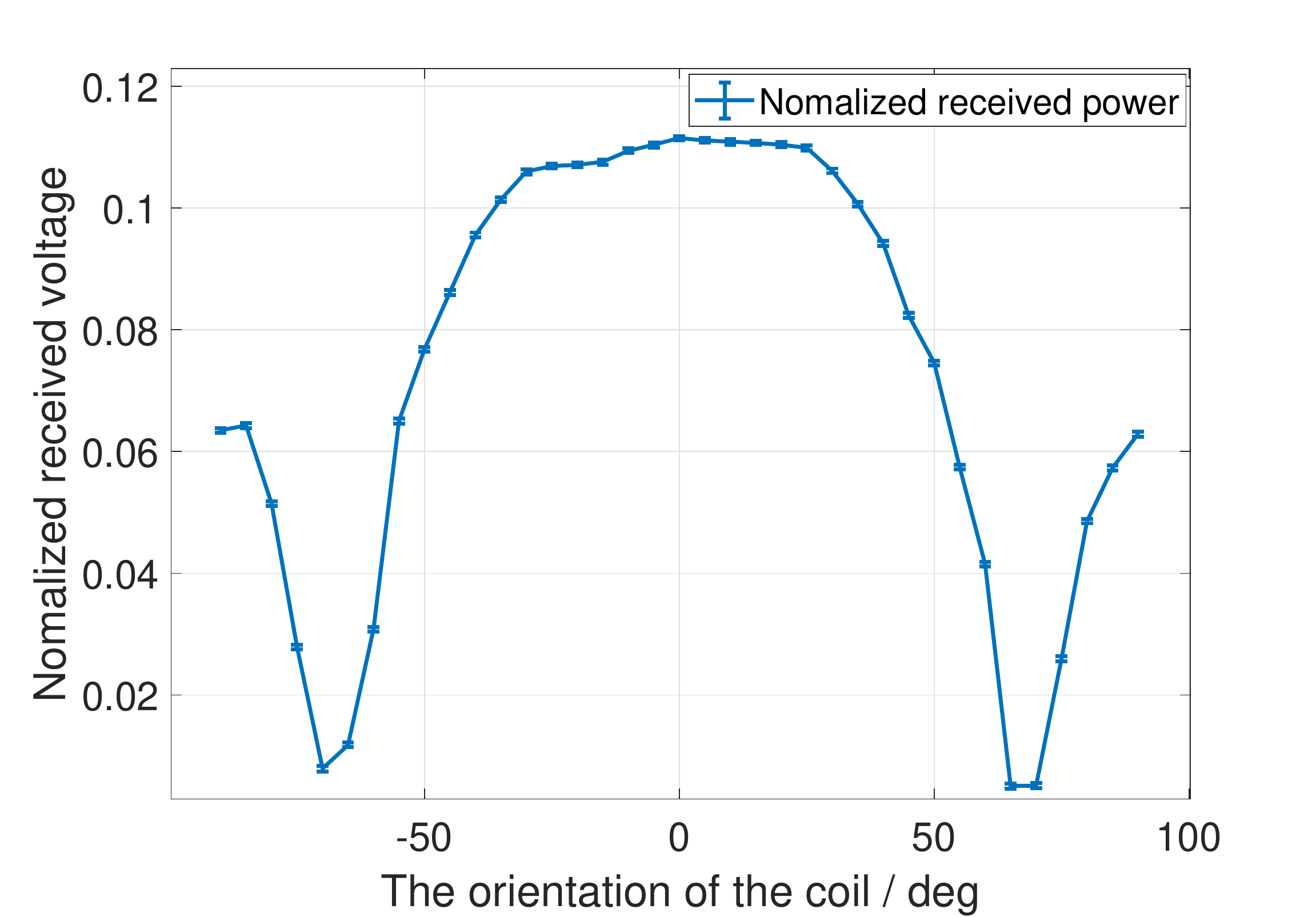}
\caption{The received voltage vs. the orientation of the coils.}
\label{orientation}
\end{center}
\end{figure}

In order to evaluate the orientation (angular misalignment) of the coils, we change the orientation of one of the two coils. Considering that we need enough space to rotate the second coil, we set a distance of 18~cm between the first coil and the second coil. When the two coils are totally parallel, we define the orientation of the second coil as 0~deg. When we rotate the second coil counterclockwise, the orientation is defined to be negative. The orientation is defined to be positive when the second coil is rotated clockwise. As Fig.~\ref{orientation} shows, the received voltage has a peak at the orientation of 0~deg. This means that we have the maximum mutual inductance at 0~deg, which validates the results in Sec.~\ref{systemdesign}. The error shown in Fig.~\ref{orientation} is within \text{$10^{-4}$}. We observe that there are two valleys in Fig.~\ref{orientation}. Since we do not consider the dynamic variability of the magnetic flux in the analysis, the second coil cuts the minimum magnetic field lines under -70~deg or 70~deg in the real experiment. Then the second coil cuts more magnetic field lines, making the energy increase when the angular misalignment is larger than 70~deg or less than -70~deg.

\subsection{The Number of coils}

\begin{figure}[htbp]
\begin{center}
\includegraphics[width=0.45\textwidth]{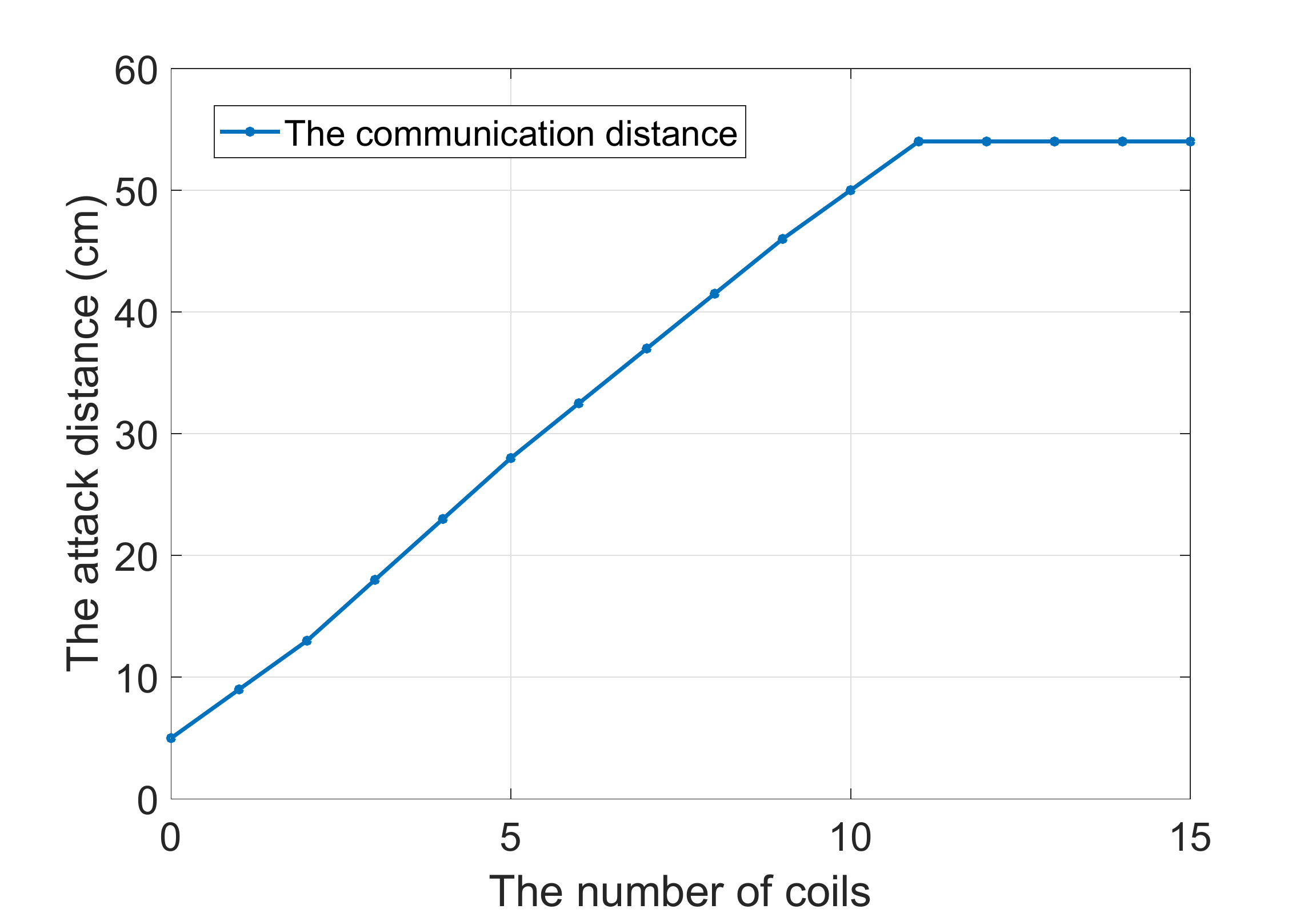}
\caption{The communication distance vs. the number of coils.}
\label{number}
\end{center}
\end{figure}

Then, we evaluate that how the number of coils impacts the attack distance when we arrange the centroid of the relay coils to be in a straight line (like dominoes). We deploy every two of the coils with a distance of 5~cm, because the NFC card receives the largest energy under the circumstances. Then, we fix a distance of 4~cm between the reader and the first relay coil, and a distance of 4~cm between the NFC card and the last coil. As Fig.~\ref{number} shows, the communication distance rises to 54~cm till we have 11 coils. We observe once the number of coils is larger than 11, the NFC card cannot wake up and respond anymore. Thus, a distance of 54~cm can be obtained when 11 coils are arranged in a straight line.
Since we obtain the best distance between two adjacent coils to maximize the output power, we keep adding the coils every 5~cm. As we know from Sec.~\ref{two-coil-system}, when we keep the maximum communication distance every time we add a new coil, the signal stops at the third coil. Therefore, we do not select the maximum communication range every time. This is the reason why the curve in Fig.~\ref{number} is almost linear before the eleventh coil is added. However, the energy is attenuated during the process of adding more coils. After we add the twelfth coil, the NFC card does not respond anymore.

\subsection{Human Body}

% The impacts of human beings on the performance of the system is evaluated. First, without the backpack, we compare the performance of the band with and without the human body. We change the distance between the DLP RFID antenna and the last square coil on the band. Fig.~\ref{humanbody1} shows the details that the farthest distance to activate the NFC tag can be 11cm. When the band is put on the human body, the received voltage is generally smaller than that without human body.

\begin{figure}[htbp]
\begin{center}
\includegraphics[width=0.45\textwidth]{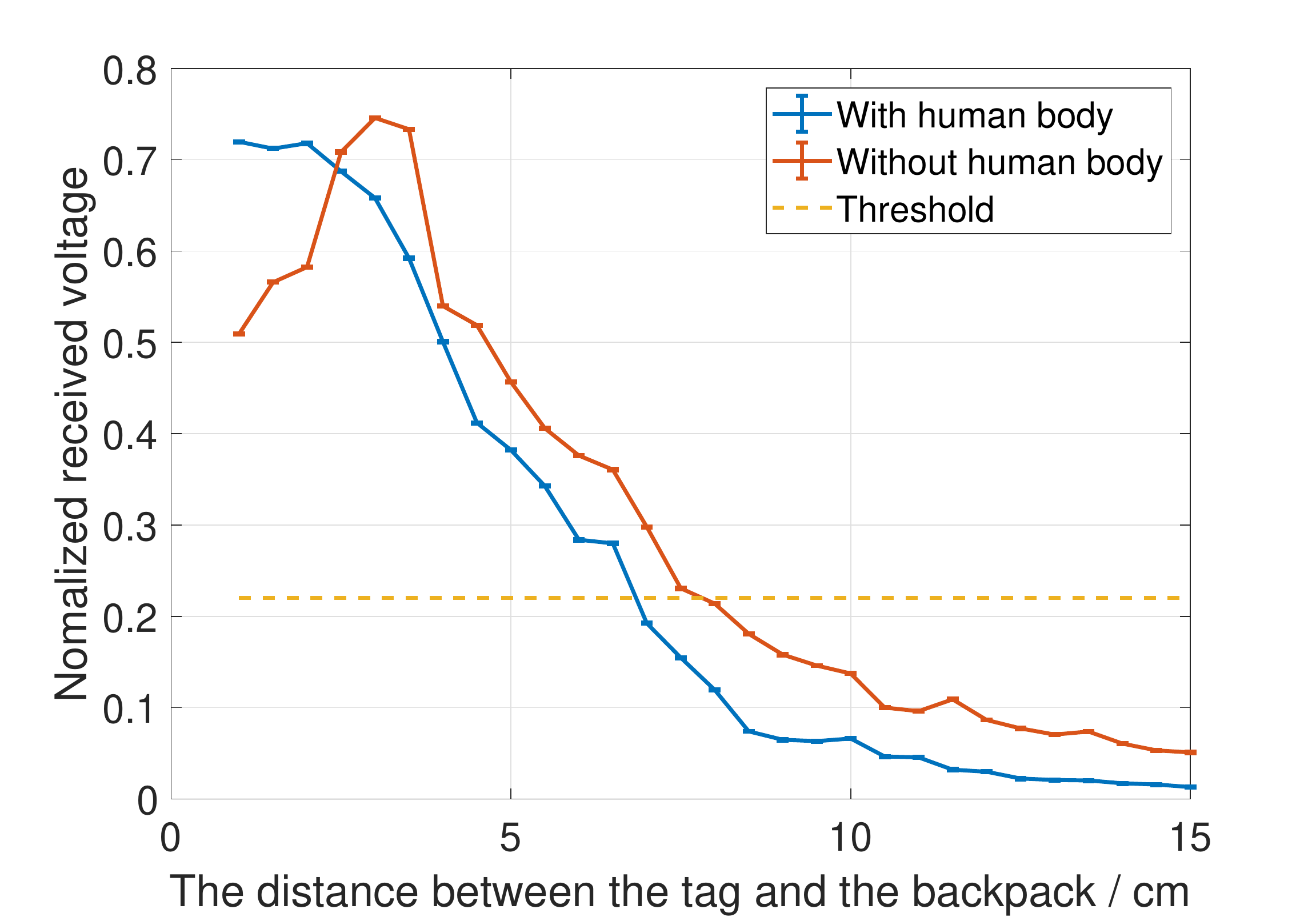}
\caption{The distance between the tag and the backpack with/without human body.}
\label{humanbody2}
\end{center}
\end{figure}

We evaluate the distance between the NFC card and the backpack when we add the three-coil box to the system, shown as Fig.~\ref{humanbody2}. We find that the distance that the device worn on human body can reach is 7~cm, while the distance of the device without human body is 8~cm. From the beginning, the energy around human body keeps dropping, this stems from the reason that the human body changes the energy transfer and as a result, changes the position of the maximum power. Then, the received voltage with human body is generally smaller than that without human body. Therefore, human bodies have the impact on the performance of the whole system, impairing the attack distance and moving the maximum point of received power. Compared to the communication range of existing NFC systems, our system can still achieve a large enlargement in range.

% \begin{figure}[htbp]
% \begin{center}
% \includegraphics[width=0.48\textwidth]{figure/bandhuman1}
% \caption{The distance between the tag and the band with/without human body.}
% \label{humanbody1}
% \end{center}
% \end{figure}

% \begin{figure}
% \centering
% \subfigure[]{
% \begin{minipage}[c]{0.48\textwidth}
% \centering
% \includegraphics[height=7.5cm,width=9cm]{figure/bandhuman1.eps}
% \end{minipage}%
% }
% \subfigure[]{
% \begin{minipage}[c]{0.48\textwidth}
% \centering
% \includegraphics[height=7.5cm,width=9cm]{figure/boxhuman1.eps}
% \end{minipage}
% }
% \caption{The distance: (a) between the tag and the band (b) between the tag and the box with/without human body.}
% \label{human}
% \end{figure}

\subsection{Impact of Multipath}

In commercial scenarios, when the attacker and the victim are standing in the same line, there would be complex multipath due to movement in the environment, like a few people passing by or other people in the line bringing mobile phones. Therefore, in order to evaluate the stability of \name , we try to simulate these situations, adding more external multipath to our system.

We first deploy some iron plates around \name\ and measure the whole distance between the reader and the NFC card. Table~\ref{interference} shows that \name\ can still has an attack distance of 48.3~cm. Then, when we place some smartphones around the band and the backpack, \name\ can still reach a communication distance of 48.2~cm. When we place a set of keys around, the whole distance between the reader and the NFC card is 48~cm. Though the metallic materials do have some impacts on \name , the results show the robustness of our system.

\begin{table}[!hpb]
\caption{The communication distance vs. different types of interference}
\label{interference}
  \centering
  \begin{tabular}
    {@{}ccccc@{}} \toprule
    %\multicolumn{2}{c}{Item} \\ \cmidrule(r){1-2}
    Interference & \tabincell{c}{No\\ Interference} & \tabincell{c}{Iron\\ plates} & Smartphones & \tabincell{c}{Keys} \\
    \midrule
    \tabincell{c}{Communication\\ distance (cm)} & 48.5 & 48.3 & 48.2 & 48 \\
 \bottomrule
  \end{tabular}
\end{table}

% \begin{figure}[htbp]
% \begin{center}
% \includegraphics[width=0.48\textwidth]{figure/multipath1}
% \caption{The whole distance with different types of interference.}
% \label{multipath}
% \end{center}
% \end{figure}

\begin{figure}[htbp]
\begin{center}
\includegraphics[width=0.45\textwidth]{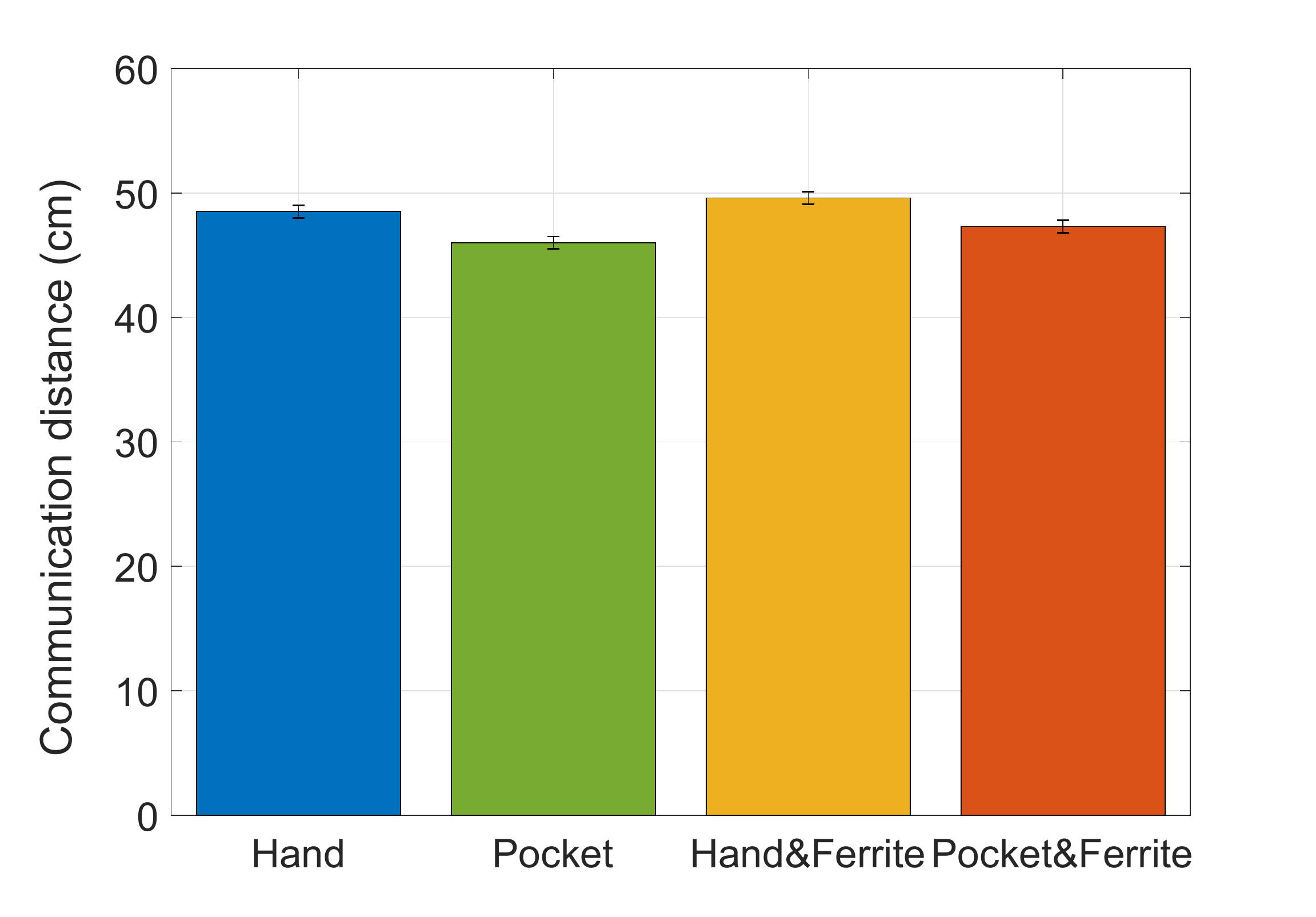}
\caption{The whole distance between the reader and the card when the card is in the hand/pocket with/without Ferrite board.}
\label{handferritebar}
\end{center}
\end{figure}

Based on the results above, we obtain the good performance on accessing the victim's NFC card. However, the people who carry some metallic objects would bring about some degree of interference and affect the power that received by the hacked cards. In order to mitigate the impact of interference from the metallic objects, we evaluate if the Ferrite board can lessen the interference. We put a piece of Ferrite board on the attacker's body before she wears the waist band on. After we use the Ferrite board, the total communication distance can be enlarged to 49.6~cm, as Fig.~\ref{handferritebar} shows. Therefore, the Ferrite material can partly reduce the interference. We also evaluate the performance when the victim puts the card into his pocket or he just holds it by hand to observe the impact of clothing. We observe that the user's clothing does have the impact on the communication distance. When the card is in the pocket, the whole distance reduces more than 1~cm. Further, the error is stable within 1~cm, showing the stability of our system.

% \begin{figure*}
% \centering
% \subfigure[]{
% \begin{minipage}[c]{0.31\textwidth}
% \centering
% \includegraphics[height=4.5cm,width=6cm]{figure/handferritebar1.eps}
% \end{minipage}%
% }
% \subfigure[]{
% \begin{minipage}[c]{0.31\textwidth}
% \centering
% \includegraphics[height=4.5cm,width=6cm]{figure/multiplecards1.eps}
% \end{minipage}
% }
% \subfigure[]{
% \begin{minipage}[c]{0.31\textwidth}
% \centering
% \includegraphics[height=4.5cm,width=6cm]{figure/successrate1.eps}
% \end{minipage}
% }
% \caption{(a) The whole distance between the reader and the card when the card is in the hand/pocket with/without ferrite sheet, (b) The whole distance between the reader and the card when there are multiple cards in the system, (c) The success rate of different card when there are multiple cards in the system.}
% \label{human}
% \end{figure*}

\begin{figure}[htbp]
\begin{center}
\includegraphics[width=0.45\textwidth]{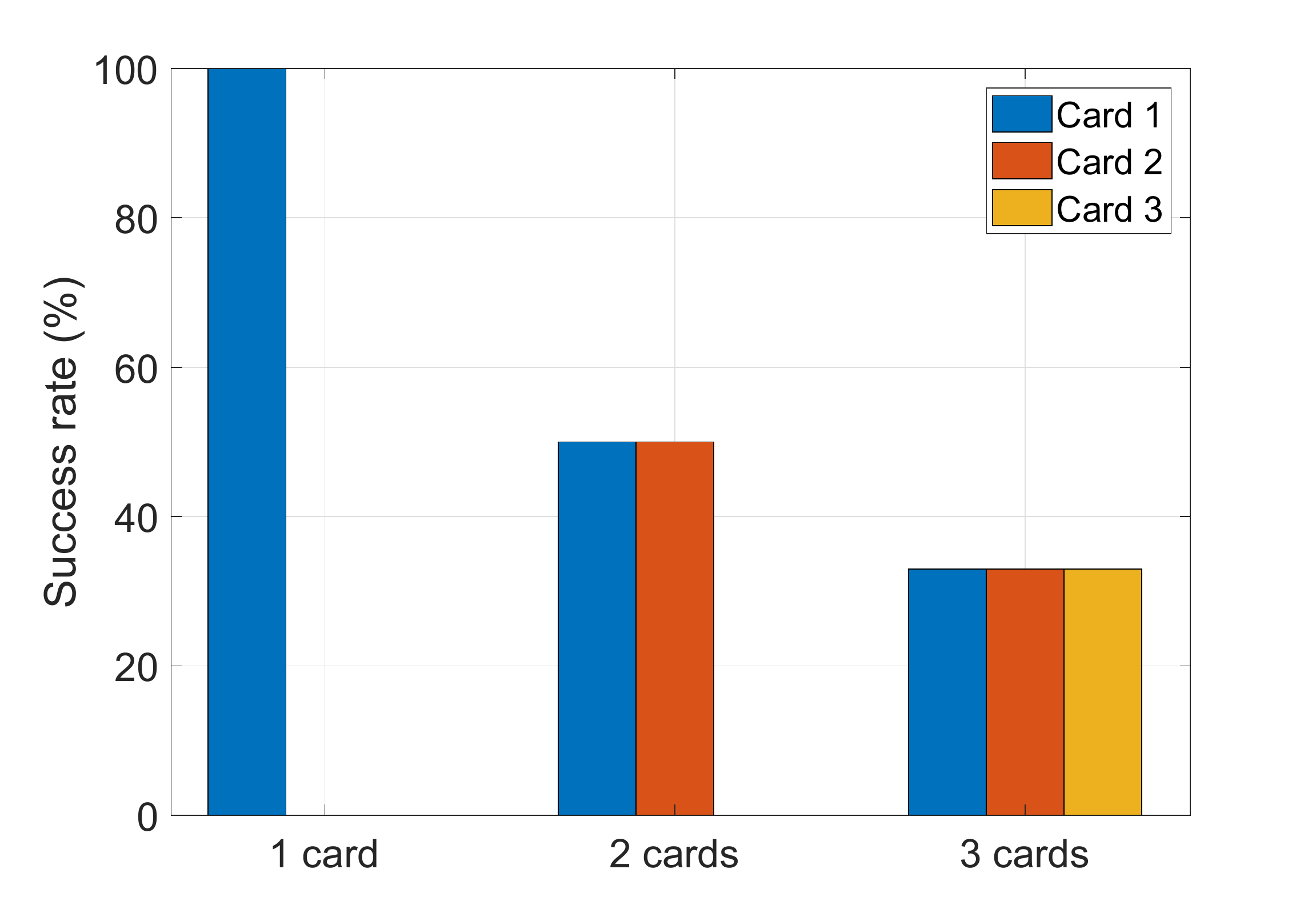}
\caption{The success rate of different card when there are multiple cards in the system.}
\label{successrate}
\end{center}
\end{figure}

We also evaluate that if our system can attack cards successfully when the victim has more than one cards. Shown in Table~\ref{multiplecards}, we find that the whole distance keeps decreasing with a stable error when the number of cards increases. This stems from the reason that a more-coil system divides the energy from the NFC reader into several parts, forcing a longer delay from the coils in response to the reader. As Fig.~\ref{successrate} shows, when the victim has two or three NFC cards, our system can still attack one of them. After extensive experiments, the results show that the three cards are randomly chosen to be attacked. In our system, if any one of the cards that the victim has is attacked, our system is regarded as hacking an NFC card successfully, i.e., \name\ can still hack cards when there are collisions happening.

% \begin{figure}[htbp]
% \begin{center}
% \includegraphics[width=0.48\textwidth]{figure/multiplecards1}
% \caption{The whole distance between the reader and the card when there are multiple cards in the system.}
% \label{multiplecards}
% \end{center}
% \end{figure}

\begin{table}[!hpb]
\caption{The communication distance vs. multiple cards}
\label{multiplecards}
  \centering
  \begin{tabular}
    {@{}cccc@{}} \toprule
    %\multicolumn{2}{c}{Item} \\ \cmidrule(r){1-2}
    \tabincell{c}{Number of\\ cards} &  one & two & three \\
    \midrule
    \tabincell{c}{Communication\\ distance (cm)} & 48.5 & 45.6 & 40.5 \\
 \bottomrule
  \end{tabular}
\end{table}

% \begin{figure}
% \centering
% \subfigure[]{
% \begin{minipage}[c]{0.45\textwidth}
% \centering
% \includegraphics[height=3.5cm,width=4.5cm]{figure/multipath1.eps}
% \end{minipage}
% }
% \subfigure[]{
% \begin{minipage}[c]{0.45\textwidth}
% \centering
% \includegraphics[height=3.5cm,width=4.5cm]{figure/smartphone1.eps}
% \end{minipage}
% }
% \caption{(a) The whole distance with different types of interference, (b) The whole distance of commercial smartphones.}
% \label{multipath}
% \end{figure}

\begin{table}[!hpb]
\caption{The communication distance vs. different smartphones}
\label{smartphone}
  \centering
  \begin{tabular}
    {@{}cccc@{}} \toprule
    %\multicolumn{2}{c}{Item} \\ \cmidrule(r){1-2}
    Smartphone & \tabincell{c}{HUAWEI\\ Mate9} &  Xiaomi6 & \tabincell{c}{SAMSUNG Galaxy \\ A9 Star} \\
    \midrule
    \tabincell{c}{Communication\\ distance (cm)} & 40 & 42 & 48 \\
 \bottomrule
  \end{tabular}
\end{table}

\subsection{Case Study}

We also use different types of commercial smartphones to act as commercial readers and read NFC cards. As Table~\ref{smartphone} shows, we use HUAWEI Mate9, Xiaomi6, SAMSUNG Galaxy A9 Star to evaluate the whole communication distance from these readers and the NFC card. The communication distance between the readers and the card in our system can still be over than 40~cm, which validates that \name\ can be effectively used in commercial scenarios.

% \begin{figure}[htbp]
% \begin{center}
% \includegraphics[width=0.48\textwidth]{figure/smartphone1}
% \caption{The whole distance of commercial smartphones.}
% \label{smartphone}
% \end{center}
% \end{figure}

\section{Discussions}\label{discussion}

Though \name\ coils are designed to be relatively inflexible, minute shape changes of the array can cause the resonant frequency to easily change, impacting the performance of the relays. In addition, our system can be affected by all-metal objects, like mobile phones or keys. Besides, the commercial readers or the passive tags should be opposite to one of the coils, ensuring as much magnetic flux as possible passes through the coil, and the tag receive the largest power. Thus in this section, we explain our limitations and future work.

\textbf{Reducing the interference.} Though the evaluation shows the robustness of our system, the system can still experience interference that absorbs the magnetic energy. Once the people around the attacker carry some metal objects, it will slightly effect the whole performance. Furthermore, when we put a smartphone face the last coil in the backpack or face the commercial reader oppositely, the NFC card behind the smartphone cannot be activated anymore. Therefore, if the victim places a mobile phone and an NFC card together when he stands in the line, our system cannot attack the card anymore. At this time, the attacker needs to find a new victim. Thus, how to relieve the interference from smartphones or how to hack the NFC-based smartphones directly would be potential problems in the future.

\textbf{Improving the robustness.}  As shown in Sec.~\ref{systemdesign} and Sec.~\ref{evaluation}, the NFC cards in our system should totally face the last coil in the backpack, because the cards can largely receive the magnetic power under the circumstances. We find that there is a tradeoff between the width and the length of the attacking area. Since we need to use the small square coil to focus the magnetic field to a certain direction, the range of attacking width is very limited. However, we have no prior knowledge on the location of the victim, weakening our success rate of attack.
% Moreover, in spite of the limited range in width of the hacking area, the attacker can still hack the NFC cards with different orientations, so long as the chip in these cards can be activated by the magnetic power. However, when the card is totally perpendicular to the last coil, our system cannot attack it anymore.
We will consider algorithms of blind search in the future to make our system more sustainable.

% Therefore, enlarging the range in width and attacking the cards with different orientations are very challenging.

\textbf{Resolving the collisions.} Though our system can attack the NFC cards when the victim has three cards or more, the attacking distance is undermined when more NFC cards are added to \name . The several cards suffer magnetic interference when they are put together, resulting in limited power for them to wake up. Collision resolution involves the optimization of anti-collision protocol, thus, using USRP to operate NFC protocols and optimizing the anti-collision mechanism should be essential to solve this problem.

% \textbf{Defence.} \name\ mainly attacks NFC cards by enlarging the communication range. When people want to combat the attacks in our system, they can put some all-mental objects, like mobile phones or some metal blocks in front of the NFC cards to impair the magnetic field.

% In the future, we will keep focusing on the design of the coils, including the release of the interference and enlargement of the range in width. In addition, in order to make them more convenient for the attacker to hack a particular NFC card, we hope to optimize the protocols based on USRP, helping the system recognize this card when collisions happen.

\section{Conclusion}\label{conclusion}

In this paper, we design the system, \name\ to attack the NFC cards by using passive relays. Unlike prior work, which required power consumption, our approach is battery-free and effortless for the attacker to wear. Instead of modifying the protocol of NFC readers, we transmit the signal of commercial NFC readers far away by passive relays, making our system stealthy and hard to be located. After designing the algorithm to search the optimal coil parameters, we build an MCR-WPT system. The waist band and the three-coil box put into the backpack can boost the communication range of NFC hacking. We attach tunable capacitors on these coils to match the unique inductance of different coils. Meanwhile, we use Ferrite board to reduce interference. Through extensive experiments, the farthest distance \name\  can achieve is 49.6~cm, which is as much as ten times the communication distance of the existing NFC system. \name\  is compatible with commercial smartphones, achieving a distance over 40~cm. Further, when the victim has several NFC cards together, \name \ can still randomly select one and attack it successfully.

\bibliographystyle{IEEEtran}
\bibliography{sample-bibliography}

% \end{IEEEbiography}


% % insert where needed to balance the two columns on the last page with
% %\newpage



%\vfill


\end{document}